\documentclass{amsart}

 \usepackage[pagewise]{lineno}

\newtheorem{theorem}{Theorem}[section]
\newtheorem{lemma}[theorem]{Lemma}
\newtheorem{construction}[theorem]{Construction}
\newtheorem{corollary}[theorem]{Corollary}
\theoremstyle{definition}
\newtheorem{definition}[theorem]{Definition}
\newtheorem{example}[theorem]{Example}

\newtheorem{proposition}[theorem]{Proposition}
\theoremstyle{remark}
\newtheorem{remark}[theorem]{Remark}

\usepackage{color,xcolor}
\usepackage{graphicx}

\newcommand{\p}{\ensuremath{{\bf Proof:\ }}}

\newcommand{\Span}{\mathrm{span}}

\begin{document}
\title[Quantum combinatorial designs and $k$-uniform states] 
{Quantum combinatorial designs and $k$-uniform states}
\author{Yajuan Zang}
\address {School of Mathematical Sciences, Hebei Normal University,
Shijiazhuang,  050024, China}
\address{School of Mathematical Sciences, Capital Normal University, Beijing, 100048, China}
\email{zyjw495@163.com}

\author{Paolo Facchi}
\address{Dipartimento di Fisica, Universit\`{a} di Bari, I-70126 Bari, Italy}
\address{INFN, Sezione di Bari, I-70126 Bari, Italy}
 \email{paolo.facchi@ba.infn.it}

\author{Zihong Tian*}
\address{School of Mathematical Sciences,
Hebei Normal University,
Shijiazhuang,  050024, China}
\email{tianzh68@163.com}


 \subjclass[2000]{ }

\thanks{{*Corresponding author: Zihong Tian}}

\begin{abstract}
Goyeneche et al.\ [Phys.\ Rev.\ A \textbf{97}, 062326 (2018)] introduced several classes of quantum combinatorial designs, namely quantum Latin squares, quantum Latin cubes, and the notion of orthogonality on them.  They also showed that mutually orthogonal quantum Latin arrangements can be entangled in the same way in which quantum states are entangled. Moreover, they established a relationship between  quantum combinatorial designs and a remarkable class of  entangled states called $k$-uniform states, i.e., multipartite pure states such that every reduction to $k$ parties is maximally mixed. In this article, we put forward the notions of incomplete quantum Latin squares and orthogonality on them and present construction methods for mutually orthogonal quantum Latin squares and mutually  orthogonal quantum Latin cubes. Furthermore, we introduce the notions of generalized mutually orthogonal quantum Latin squares and generalized  mutually orthogonal quantum Latin cubes, which are equivalent to quantum orthogonal arrays of size $d^2$ and $d^3$, respectively, and thus naturally provide $2$- and $3$-uniform states.

\end{abstract}

\maketitle

\section{Introduction}

Entanglement is considered to be one of the most striking features of quantum mechanics and has been widely utilized as a crucial resource in quantum information science~\cite{Chuang,Benenti}, from quantum computation~\cite{Jozsa} to  quantum teleportation~\cite{BennettC2}  and quantum key distribution~\cite{BennettC1,Lo}. The research on multipartite entanglement is
no simple matter. Recently, a striking class of $N$-party entangled pure states, called $k$-uniform states, have attracted much
attention. These states have the property that every reduction to $k$ parties is maximally mixed, where $k\leq \lfloor N/2 \rfloor$, with $\lfloor.\rfloor$ denoting the floor function~\cite{Goyeneche0}. When $k=\lfloor N/2 \rfloor$, these states, known as maximally multipartite entangled states~\cite{Facchi2}, or absolutely maximally entangled  (AME) states~\cite{Helwig}, exhibit maximal entanglement in all possible partitions and thus play a pivotal role in quantum secret sharing, multipartite teleportation, and in tensor network states for holographic codes~\cite{Zhang,Latorre}.

So far, plenty of work has been done for finding the application and the existence of $k$-uniform states~\cite{Facchi1,Facchi2,Helwig}.  Orthogonal array is a very important configuration in combinatorial design. Recently, Goyeneche and \.{Z}yczkowski  established a link between a special kind of orthogonal arrays and $k$-uniform states~\cite{Goyeneche0}. Moreover, Zang, Li and Pang et al.\ presented some 2, 3-uniform states by those orthogonal arrays~\cite{Li,Pang1,Zang,Zang1}.
Besides, Latin square (LS) is another significative configuration in combinatorial design and has a long history~\cite{Donald}. Latin squares have wide applications in many fields ranging from quantum information, to experimental designs and cryptology. In particular, orthogonal Latin squares have a very closed connection with mutually unbiased bases ~\cite{Hayashi,Paterek0,Paterek,Song}.

In recent years, Musto and Vicary  introduced the notions of quantum Latin square (QLS)~\cite{Musto1}, weakly orthogonal QLSs  and  orthogonal QLSs ~\cite{Musto3}, where classical symbols appearing in entries of arrangements were extended to quantum states. These concepts could be used to construct unitary error bases and mutually unbiased bases~\cite{Musto1,Musto2,Musto3}. In 2018, Goyeneche et al.\ put forward the concept of quantum Latin cube (QLC) and quantum Latin hypercube ~\cite{Goyeneche1}. They also introduced the notions of orthogonal QLCs and  orthogonal quantum Latin hypercubes. Moreover, they identified a crucial ingredient missing in the previous approach in~\cite{Musto3}: they  pointed out that a pair of orthogonal QLSs could be entangled in such a way that they cannot be expressed as two separated arrangements, the same with orthogonal QLCs and orthogonal quantum Latin hypercubes. These entangled designs are intrinsically associated with quantum orthogonal arrays (QOAs)~\cite{Goyeneche1}, which can generate $k$-uniform states.

A self-orthogonal Latin square (SOLS) is a special kind of orthogonal LSs, which is orthogonal to its transpose, thus it is not equivalent with a pair of orthogonal LSs. Indeed, SOLS takes up less storage space in experimental designs than orthogonal LSs, which is one of the reasons why it is an interesting concept in combinatorial designs. In this article, we will introduce a quantum version of SOLS, which will be named self-orthogonal quantum Latin square (SOQLS). Primarily we will exhibit construction methods of  mutually orthogonal quantum Latin squares (MOQLSs), mutually  orthogonal quantum Latin cubes (MOQLCs), such that families of $k$-uniform states can be obtained, with $k=2,3$. Furthermore, we will introduce  generalizations of MOQLSs and MOQLCs in which the arrangements may be entangled, so that they  will have a one-to-one relationship with QOAs.

The article is organized as follows. In Section~\ref{sec:QLS} we present two construction methods: One is the direct product for MOQLSs and the other is filling in holes for SOQLS. Interestingly enough, the obtained MOQLSs and SOQLS are not equivalent with each other. Meanwhile, we define the notions of incomplete quantum Latin squares (IQLSs) and orthogonality on them as tools for the construction of filling in holes.  In Section~\ref{sec:QLC} we give a notion of mutually orthogonal quantum Latin cubes (MOQLCs), which is different from the one in~\cite{Goyeneche1}. Moreover, we show a construction method of direct product for MOQLCs. In Section~\ref{sec:GQLS} we introduce the notions of generalized mutually orthogonal quantum Latin squares (GMOQLSs) and generalized  mutually orthogonal quantum Latin cubes (GMOQLCs), whose arrangements may be entangled. Actually, MOQLSs and MOQLCs are  special cases of GMOQLSs and GMOQLCs when the arrangements are fully separated.  Moreover we give direct proofs of the one-to-one relationships between GMOQLSs and QOAs, as well as GMOQLCs and QOAs with size $d^2$ and $d^3$, respectively.
 Finally, after setting up the quantum combinatorial designs, we  get a family of $k$-uniform states and absolutely maximally entangled states.
In Section~\ref{sec:concl} we gather and discuss the main results obtained in this article and draw our conclusions.

\section{Quantum Latin squares}
\label{sec:QLS}

\subsection{Classical Latin squares}

In this section, we review some basic combinatorial concepts used in this work.
A (classical) \emph{Latin square} of order $d$ denoted by LS$(d)$ is a $d \times d$ square in which each of the numbers $0,1,\dots, d - 1$ occurs exactly once
in each row and exactly once in each column. Two Latin squares $L_1,L_2$ of order $d$ are \emph{orthogonal}, if when $L_1$ is superimposed on $L_2$, every ordered pair $00, 01, \dots,d - 1\, d - 1$ occurs. A set of $t\geq2$ \emph{mutually orthogonal Latin squares} of order $d$, denoted by $t$-MOLS$(d)$, is a set of Latin squares $L_1,\dots,L_t$ $(t\geq 2)$ such that
every $i,j$, $1 \leq i < j\leq t$, $L_i$ and $L_j$ are orthogonal.
A \emph{self-orthogonal} Latin square (SOLS) is a Latin square that is orthogonal to its transpose. The reader can see the references~\cite{Brayton,Hedayat,Colbourn,Donald} for deep research on them.

\begin{lemma}\label{SOLS}(\cite{Brayton})
There exists a SOLS$(d)$ if and only if $d\geq 4$ and $d \neq 6$.
\end{lemma}
Let $m(d)$ be the largest number of mutually orthogonal classical Latin squares of order $d$.

\begin{lemma}(\cite{Colbourn})\label{m(d)}
For any integer $d\geq 2$, $m(d)\leq d-1$.
\end{lemma}

An \emph{orthogonal array} of size $r$, with $N$ factors, $d$ levels, and strength $k$, denoted by OA$(r,N,d,k)$,
is an $r \times N$ array $A$ over a set $S$ of $d$ symbols such that every $r\times k$ subarray contains
each $k$-tuple based on $S$ exactly $\lambda$ times as a row, where $\lambda=r/d^k$~\cite{Hedayat}.

Actually, mutually orthogonal classical Latin squares have an equivalence relationship with orthogonal arrays of strength 2 and $\lambda=1$.

\begin{lemma}(\cite{Hedayat})\label{OA2}
There exists a $t$-MOLS$(d)$  if and only if there exists an OA$(d^2,t+2,d,2)$.
\end{lemma}

As a consequence of the relation between MOLSs and OAs given by Lemma~\ref{OA2}, there are some results about the largest number of MOLSs $m(d)$.
\begin{lemma}\label{md}(\cite{Bose,Bush,Colbourn,Hedayat})

	1)  If $q$ is a prime power, then $m(q)=q-1$.

	2) Suppose that $d = p_{1}^{r_1}p_{2}^{r_2}\ldots p_{s}^{r_s} $, where $s\geq 2$, $r_i$ is a positive integer, $p_i$  is a prime and $p_i\neq p_j$ for $1\leq i\neq j\leq s$,   then $m(d)\geq min\{p_{i}^{r_i}-1 : 1 \leq i \leq s\}$.

	3) For any $d\neq 2,6$, $m(d)\geq 2$.

	4) For any $d\neq 2,3,6,10$, $m(d)\geq 3$.

	5) For any $d\neq 2,3,4,6,10,22$, $m(d)\geq 4$.

\end{lemma}

\subsection{Quantum Latin squares}

Recently, quantum Latin square (QLS)~\cite{Musto1} and orthogonal QLSs~\cite{Musto3} have been introduced. In this section, we review the concepts of QLS and orthogonal QLSs, but also generalize the orthogonality of two QLSs to $t$  QLSs, and self-orthogonal quantum Latin square.
In the following,  let $[d]=\{0,1,\dots,d-1\}$  and $S_{d}$ be the symmetric group  on the set  $[d]$ .
\begin{definition} \label{QLS}
 A quantum Latin square  $\Phi$ of dimension $d$ denoted by QLS$(d)$ is a $d\times d$ array of vectors $|\Phi_{i,j}\rangle\in \mathbb{C}^{d}$, $i,j\in [d]$,
such that every row and every column determine an orthonormal basis of the complex vector space $\mathbb{C}^{d}$.
\end{definition}

Two classical Latin squares are said to be equivalent if one can be transformed into the other by
permutations of the rows, columns or relabeling the symbols. Similarly, there is a notion of
equivalence between two quantum Latin squares~\cite{Musto1}.

\begin{definition}
Two quantum Latin squares $\Phi$, $\Psi$ of dimension $d$ are equivalent if there exists a
unitary operator $U$ on $\mathbb{C}^d$, a set of modulus-1 complex numbers $c_{ij}$, and two permutations $\sigma,\tau \in S_d$, such
that the following holds for all $i, j \in [d]$:
\begin{equation}
|\Psi_{i,j}\rangle = c_{ij}U|\Phi_{\sigma(i),\tau(j)}\rangle.
\end{equation}
\end{definition}

By associating with each number $l\in [d]$ in a classical Latin square of order $d$ and the computational basis element $|l\rangle \in \mathbb{C}^d$, we  get a quantum Latin square for which the elements in every row or column form a
computational basis, and we call it a \emph{classical quantum Latin square}.
Moreover, if a quantum Latin square is equivalent to a classical one, then we also call it a classical quantum Latin square, otherwise, it is a \emph{non-classical quantum Latin square} \cite{Musto2} or a \emph{genuinely quantum Latin square} \cite{Paczos}.

\begin{lemma}\label{classical}
If~ $\Phi$ is a classical quantum Latin square of dimension $d$, then for any~$i,j,m,n\in [d]$, it satisfies ~$|\langle\Phi_{i,j}|\Phi_{m,n}\rangle|=0$ or~1.
\end{lemma}
\noindent \p Let $l=(l_{i,j})$ be a classical Latin square of order $d$, and ~$L=(|l_{i,j}\rangle)$ be the corresponding  classical quantum Latin square of $l$.  Then for any $i,j,m,n\in [d]$, it should be true that $\langle l_{i,j}|l_{m,n}\rangle=0$ or~1. Suppose $\Phi$ is equivalent to~$L$, then there exists a
unitary operator $U$ on $\mathbb{C}^d$, a family of modulus-1 complex numbers $c_{ij}$, and two permutations $\sigma,\tau \in S_d$, such that for any~ $i,j\in [d]$, the equation ~$|l_{i,j}\rangle= c_{ij}U|\Phi_{\sigma(i),\tau(j)}\rangle$ holds. Thus, $\langle l_{i,j}|l_{m,n}\rangle=c^*_{ij}c_{m,n}\langle\Phi_{\sigma(i),\tau(j)}|U^{\dagger}U|\Phi_{\sigma(m),\tau(n)}\rangle=c^*_{ij}c_{m,n}$\\$\langle\Phi_{\sigma(i),\tau(j)}|\Phi_{\sigma(m),\tau(n)}\rangle=0$ or~1.  Since~$c^*_{ij}$, $c_{m,n}$ are   modulus-1 complex numbers, and~$\sigma,\tau \in S_d$, then for any ~$i,j,m,n\in [d]$, it is true that~$|\langle\Phi_{i,j}|\Phi_{m,n}\rangle|=0$ or~1.
\qed

\begin{definition}\label{MOQLS1}
Two quantum Latin squares $\Phi$,$\Psi$ of dimension $d$ are \emph{orthogonal} if the set of vectors
$\{|\Phi_{i,j}\rangle\otimes |\Psi_{i,j}\rangle: i,j\in [d ] \} $ forms an orthonormal basis of the space $\mathbb{C}^{d}\otimes \mathbb{C}^{d}$,
i.e., $\langle\Phi_{i,j}\otimes \Psi_{i,j}\, |\,\Phi_{i',j'}\otimes \Psi_{i',j'}\rangle=
\langle\Phi_{i,j}|\Phi_{i',j'}\rangle\langle\Psi_{i,j}|\Psi_{i',j'}\rangle=\delta_{ii'}\delta_{jj'}$,
for $i,j,i',j'\in [d]$.
\end{definition}

The orthogonality of quantum Latin squares is unaffected by conjugation of one of the squares~\cite{Musto2}.
\begin{definition}
 Given a quantum Latin square $\Phi$, its conjugate $\Phi^*$, is the quantum Latin square with entries
$(|\Phi^*_{i,j}\rangle) = (|\Phi_{i,j}\rangle^*)$ for $i,j\in [d] $.
\end{definition}
\begin{lemma}(\cite{Musto2})\label{conjugate}
Two quantum Latin squares $\Phi$, $\Psi$ are orthogonal if and only if $\Phi^*$, $\Psi$ are orthogonal.
\end{lemma}

Similar with the concept of self-orthogonal (classical) Latin square, we give a definition of self-orthogonal quantum Latin square.

\begin{definition}
Given a quantum Latin square $\Phi$, its transpose $\Phi^{\mathrm{T}}$ is the quantum Latin square with entries $(|\Phi^{\mathrm{T}}_{i,j}\rangle)=(|\Phi_{j,i}\rangle)$ for $i,j\in [d]$.
\end{definition}

\begin{definition}
Given a quantum Latin square $\Phi$, its conjugate transpose $\Phi^{\dag}$ is the quantum Latin square with entries $(|\Phi^{\dag}_{i,j}\rangle)=(|\Phi_{j,i}\rangle^{*})$ for $i,j\in [d] $.
\end{definition}
\begin{definition}
Let $\Phi$ be a quantum Latin square of dimension ~$d$. If~$\Phi$ is orthogonal to its conjugate transpose, then we call it a  self-orthogonal quantum Latin square, and denote it by~SOQLS$(d)$.
\end{definition}

From  Lemma~\ref{conjugate}, we know that $\Phi$ is orthogonal to its conjugate transpose $\Phi^{\dag}$ if and only if  $\Phi$ is orthogonal to its transpose $\Phi^{\mathrm{T}}$. So we have the following lemma.
\begin{lemma}\label{soqls2}
 $\Phi$ is a SOQLS$(d)$ if and only if $\Phi$ is orthogonal to its transpose $\Phi^{\mathrm{T}}$.
\end{lemma}

\begin{lemma}
If $\Phi$ is a SOQLS$(d)$, then $d\geq 4$; moreover, $\{|\Phi_{ii}\rangle: i\in [d]\}$ forms an orthonormal basis of the space $\mathbb{C}^{d}$.
\end{lemma}
\noindent \p Since for any $i,j\in[d]$, $\langle\Phi_{jj}\otimes\Phi_{jj}|\Phi_{ii}\otimes \Phi_{ii}\rangle=\langle\Phi_{jj}|\Phi_{ii}\rangle^2=\delta_{ij}$. Thus $\{|\Phi_{ii}\rangle: i\in [d]\}$ forms an orthonormal basis of the space $\mathbb{C}^{d}$. The impossibility of $d=2$ is obvious. If a SOQLS$(3)$ exists, then $\{|\Phi_{ii}\rangle: i\in [3]\}$ forms an orthonormal basis of the space $\mathbb{C}^{3}$. Furthermore $\langle\Phi_{01}|\Phi_{ii}\rangle=0$ and $\langle\Phi_{10}|\Phi_{ii}\rangle=0$ for $i=0,1$,  which is in contradiction with $\langle\Phi_{01}\otimes\Phi_{10}|\Phi_{22}\otimes\Phi_{22}\rangle=0$. Therefore $d\geq 4$. \qed

\begin{example} (Non-classical SOQLS)\label{ex1}
There exists a SOQLS(14).
\end{example}
Let {\small $$|\phi_1\rangle= \frac{ |10\rangle+|11\rangle+|12\rangle+|13\rangle}{2}, \qquad |\phi_2\rangle= \frac{ |10\rangle-|11\rangle+|12\rangle-|13\rangle}{2},$$}
{\small$$|\phi_3\rangle= \frac{ |10\rangle+|11\rangle-|12\rangle-|13\rangle}{2}, \qquad|\phi_4\rangle= \frac{ |10\rangle-|11\rangle-|12\rangle+|13\rangle}{2}.$$}

Then,
$${\small \begin{array}{lc}
\setlength{\arraycolsep}{1.0 pt}
\mbox{}&
\hspace{-0.6cm}\begin{array}{|c|c|c|c|c|c|c|c|c|c|c|c|c|c|}
\hline
|0\rangle & |6\rangle &|13\rangle& |7\rangle &|12\rangle &  |3\rangle  &|8\rangle& |10\rangle& |9\rangle &|11\rangle &  |5\rangle  &|4\rangle& |2\rangle& |1\rangle \\
\hline
|10\rangle & |1\rangle &|7\rangle& |12\rangle &|5\rangle &  |11\rangle  &|2\rangle& |4\rangle& |13\rangle &|3\rangle &  |9\rangle  &|6\rangle& |8\rangle& |0\rangle \\
\hline
|8\rangle & |11\rangle &|2\rangle& |9\rangle &|7\rangle &  |13\rangle  &|10\rangle& |6\rangle& |12\rangle &|1\rangle &  |4\rangle  &|5\rangle& |0\rangle& |3\rangle \\
\hline
|13\rangle & |7\rangle &|10\rangle& |3\rangle &|6\rangle &  |4\rangle  &|9\rangle& |1\rangle& |11\rangle &|12\rangle &  |8\rangle  &|0\rangle& |5\rangle& |2\rangle \\
\hline
|9\rangle & |12\rangle &|0\rangle& |11\rangle &|4\rangle &  |6\rangle  &|3\rangle& |2\rangle& |10\rangle &|13\rangle &  |7\rangle  &|8\rangle& |1\rangle& |5\rangle \\
\hline
|6\rangle & |8\rangle &|1\rangle& |10\rangle &|13\rangle &  |5\rangle  &|12\rangle& |11\rangle& |7\rangle &|2\rangle &  |0\rangle  &|3\rangle& |9\rangle& |4\rangle \\
\hline
|12\rangle & |9\rangle &|8\rangle& |13\rangle &|11\rangle &  |0\rangle  &|6\rangle& |5\rangle& |3\rangle &|10\rangle &  |2\rangle  &|1\rangle& |4\rangle& |7\rangle \\
\hline
|5\rangle & |13\rangle &|12\rangle& |8\rangle &|10\rangle &  |2\rangle  &|11\rangle& |7\rangle& |4\rangle &|0\rangle &  |1\rangle  &|9\rangle& |3\rangle& |6\rangle \\
\hline
|11\rangle & |5\rangle &|3\rangle& |0\rangle &|1\rangle &  |10\rangle  &|13\rangle& |12\rangle& |8\rangle &|4\rangle &  |6\rangle  &|2\rangle& |7\rangle& |9\rangle \\
\hline
|4\rangle & |10\rangle &|11\rangle& |1\rangle &|2\rangle &  |12\rangle  &|0\rangle& |13\rangle& |5\rangle &|9\rangle &  |3\rangle  &|7\rangle& |6\rangle& |8\rangle \\
\hline
|7\rangle & |0\rangle &|6\rangle& |2\rangle &|9\rangle &  |8\rangle  &|4\rangle& |3\rangle& |1\rangle &|5\rangle &|\phi_1\rangle  &|\phi_2\rangle& |\phi_3\rangle& |\phi_4\rangle \\
\hline
|1\rangle & |2\rangle &|9\rangle& |4\rangle &|3\rangle &  |7\rangle  &|5\rangle& |8\rangle& |0\rangle &|6\rangle &  |\phi_4\rangle &|\phi_3\rangle& |\phi_2\rangle& |\phi_1\rangle \\
\hline
|3\rangle & |4\rangle &|5\rangle& |6\rangle &|0\rangle &  |1\rangle  &|7\rangle& |9\rangle& |2\rangle &|8\rangle &  |\phi_2\rangle &|\phi_1\rangle& |\phi_4\rangle&|\phi_3\rangle \\
\hline
|2\rangle & |3\rangle &|4\rangle& |5\rangle &|8\rangle &  |9\rangle  &|1\rangle& |0\rangle& |6\rangle &|7\rangle & |\phi_3\rangle&|\phi_4\rangle&|\phi_1\rangle& |\phi_2\rangle \\
\hline
\end{array}
\end{array}}$$
is a non-classical SOQLS(14).

\vspace{0.4cm}
A set of $t\geq 2$ quantum Latin squares of dimension $d$, say $\Phi_1, \Phi_2,\dots, \Phi_t$, is said to be \emph{mutually
orthogonal}, and is denoted by $t$-MOQLS$(d)$,
if $\Phi_i$ and $\Phi_j$  are orthogonal for all $1 \leq i < j \leq t$.

 Let $M(d)$ be the largest number of mutually  orthogonal  non-classical quantum Latin squares of dimension $d$. Analogously to classical Latin squares, an upper bound to $M(d)$ can be proved.
 \begin{lemma}(\cite{Musto2})\label{Md0}
For any integer $d\geq 2$,
$M(d)\leq d-1$.
\end{lemma}

In the following, we will focus  on the bound which can be reached for  mutually  orthogonal  non-classical quantum Latin squares.

\subsection{Direct product construction}
\label{sec:directproduct}
In this subsection, we will provide a construction of mutually orthogonal quantum Latin squares by direct product. In particular, we describe a method to construct  mutually  orthogonal non-classical quantum Latin squares from the mutually  orthogonal  classical Latin squares.

Let $V$ and $W$ be Hilbert spaces of dimension $d_1$ and $d_2$ respectively.
Then the tensor product $V \otimes W$ is a Hilbert space of dimension $d_1d_2$, whose elements are linear combinations of `tensor products' $|v\rangle\otimes|w\rangle$ of elements $|v\rangle$ of $V$ and $|w\rangle$ of $W$. In particular, if $\{|i\rangle\}$ and $\{|j\rangle\}$ are orthonormal basis of the spaces $V$ and $W$, respectively, then $\{|i\rangle\otimes |j\rangle\}$ is an orthonormal basis of $V \otimes W$, whence $\mathbb{C}^{d_1d_2}\simeq\mathbb{C}^{d_{1}}\otimes \mathbb{C}^{d_{2}}$~\cite{Chuang}.

\begin{construction}\label{qod1d2}(Direct Product Construction)
If there exists a  2-MOQLS$(d_1)$ and a  2-MOQLS$(d_2)$, then there exists a  2-MOQLS$(d_1d_2)$.
\end{construction}
\noindent \p Suppose $\Phi^1=(|\Phi^1_{i,j}\rangle)$, $\Phi^2=(|\Phi^2_{i,j}\rangle)$ is a pair of orthogonal quantum Latin squares of dimension $d_1$, and $\Psi^1=(|\Psi^1_{m,n}\rangle)$, $\Psi^2=(|\Psi^2_{m,n}\rangle)$ is a pair of orthogonal quantum Latin squares of dimension $d_2$. Then  $\Phi=(|\Phi_{(i,m),(j,n)}\rangle)=\Phi^1\otimes \Psi^1$ and $\Psi=(|\Psi_{(i,m),(j,n)}\rangle)=\Phi^2\otimes \Psi^2$ is a pair of  orthogonal quantum Latin squares of dimension $d_1d_2$,
where $|\Phi_{(i,m),(j,n)}\rangle=|\Phi^1_{i,j}\rangle\otimes|\Psi^1_{m,n}\rangle$ and $|\Psi_{(i,m),(j,n)}\rangle=|\Phi^2_{i,j}\rangle\otimes|\Psi^2_{m,n}\rangle$.

In fact,  the set of vectors
$\{|\Phi_{(i,m),(j,n)}\rangle\otimes |\Psi_{(i,m),(j,n)}\rangle: i,j\in [d_1],m,n\in [d_2] \} $ forms an orthonormal basis of the space $\mathbb{C}^{d_1d_2}\otimes \mathbb{C}^{d_1d_2}$. Indeed,
\begin{eqnarray*}
\hspace{-3cm}&&\hspace{-0.5cm}(|\Phi_{(i,m),(j,n)}\rangle\otimes |\Psi_{(i,m),(j,n)}\rangle,|\Phi_{(i',m'),(j',n')}\rangle\otimes |\Psi_{(i',m'),(j',n')}\rangle)\\
	 &&\hspace{-0.5cm}=((|\Phi^1_{i,j}\rangle\otimes|\Psi^1_{m,n}\rangle)\otimes (|\Phi^2_{i,j}\rangle\otimes|\Psi^2_{m,n}\rangle), (|\Phi^1_{i',j'}\rangle\otimes|\Psi^1_{m',n'}\rangle)\otimes (|\Phi^2_{i',j'}\rangle\otimes|\Psi^2_{m',n'}\rangle))\\
 &&\hspace{-0.5cm}=(|\Phi^1_{i,j}\rangle\otimes|\Psi^1_{m,n}\rangle,|\Phi^1_{i',j'}\rangle\otimes|\Psi^1_{m',n'}\rangle)
(|\Phi^2_{i,j}\rangle\otimes|\Psi^2_{m,n}\rangle,|\Phi^2_{i',j'}\rangle\otimes|\Psi^2_{m',n'}\rangle)\\
 &&\hspace{-0.5cm}=\langle\Phi^1_{i,j}|\Phi^1_{i',j'}\rangle\langle\Phi^2_{i,j}|\Phi^2_{i',j'}\rangle
\langle\Psi^1_{m,n}|\Psi^1_{m',n'}\rangle\langle\Psi^2_{m,n}|\Psi^2_{m',n'}\rangle\\
 &&\hspace{-0.5cm}=\delta_{ii'}\delta_{jj'}\delta_{mm'}\delta_{nn'}.
\end{eqnarray*}
\qed

The construction can be easily generalized to $t$ mutually orthogonal quantum Latin squares.
\begin{corollary}\label{mqod1d2}
Let $l\geq 2$. If there exist a $t_j$-MOQLS$(d_j)$, for any $1\leq j\leq l$,  then there exists  a $t$-MOQLS$(d)$, where $t= \min\{t_1,t_2,\dots,t_l\}$ and $d=d_1d_2\cdots d_l$.
\end{corollary}

In particular, mutually orthogonal quantum Latin squares of dimension $d_1d_2$ can be established from  mutually orthogonal classical quantum Latin squares of dimension $d_1$ and $d_2$ after the action of unitary matrices.

\begin{construction}\label{d1d2}
If there exists a  2-MOLS$(d_1)$ and a 2-MOLS$(d_2)$, then there exists a  2-MOQLS$(d_1d_2)$.
\end{construction}

See Appendix~\ref{sec:AppA} for the proof of Construction~\ref{d1d2}. Analogously to Corollary~\ref{mqod1d2}, we  can generalize the result as follows.

\begin{corollary}\label{4.14}
Let $l\geq 2$ and $d=d_1d_2\cdots d_{l}$, with $ m(d_j)\geq 2$ for all $1\leq j\leq l$. Then there exists a $t$-MOQLS$(d)$ with $t=\min\{m(d_1),m(d_2),\dots, m(d_l)\}$.
\end{corollary}

From the proof of Construction~\ref{d1d2}, for given suitable unitary matrices we  get plenty of non-classical quantum Latin squares by different choices of the $\tau$s in each block of $\Phi$ or $\Psi$. Actually, for different choice of  $\tau$ in each block of the two squares, we can get different 2-MOQLS$(d_1d_2)$s. Obviously, we cannot choose $\tau$s all being $\mathbb{I}$ or $U$ in  $\Phi$ or $\Psi$, if we want to get non-classical quantum Latin squares.

\begin{example} (Non-classical 2-MOQLSs)\label{ex3}
There exists a  2-MOQLS(12).
\end{example}
\noindent \p Let $\mathbb{C}^3= \Span  \{|0\rangle,|1\rangle,|2\rangle\}$ and $\mathbb{C}^4= \Span\{|0\rangle,|1\rangle,|2\rangle,|3\rangle\}$. Then $\mathbb{C}^{12}\simeq \mathbb{C}^4\otimes \mathbb{C}^3=\Span\{|i\rangle \otimes |j\rangle: i \in [4] , j \in [3] \}=\Span\{|0\rangle,|1\rangle,\dots,|11\rangle\}$. Define $U=\sum\limits_{i\in[4]}|i\rangle\langle i|\otimes U_i$,
where
 \vspace{-0.4cm}\begin{equation*}
\begin{small}
\setlength{\arraycolsep}{2pt}
 U_0=\left(
       \begin{array}{ccc}
1 & 1& 1\\
1 & e^{\frac{2\pi \sqrt{-1}}{3}}& e^{\frac{-2\pi \sqrt{-1}}{3}}\\
1 & e^{\frac{-2\pi \sqrt{-1}}{3}}& e^{\frac{2\pi \sqrt{-1}}{3}}\\
\end{array}
        \right), \hspace{0.2 cm}
        U_1=\frac{1}{\sqrt{3}}\left(
        \begin{array}{ccc}
1+\sqrt{-1} & \frac{1-\sqrt{-1}}{\sqrt{2}}& 0\\
-\sqrt{\frac{-1}{2}} & 1 & \frac{1}{\sqrt{2}}+\sqrt{-1}\\
\frac{1}{\sqrt{2}} & \sqrt{-1}&1-\sqrt{\frac{-1}{2}} \\
\end{array}
        \right),\end{small}\end{equation*}
  \vspace{0cm} \begin{equation*}
  \begin{small}
\setlength{\arraycolsep}{5pt}
U_2=\left(
        \begin{array}{ccc}
\frac{1}{\sqrt{2}} &\frac{1}{\sqrt{2}}  & 0\\
\frac{1}{\sqrt{2}}  & -\frac{1}{\sqrt{2}}  & 0\\
0 & 0 & \frac{1}{\sqrt{2}}  \\
\end{array}
        \right),  \vspace{0.2 cm} \hspace{0.3 cm}~~~~~
        U_3=\left(
       \begin{array}{ccc}
\frac{2}{3} & \frac{2}{3}& \frac{1}{3}\\
\frac{1}{3}& -\frac{2}{3}&\frac{2}{3}\\
-\frac{2}{3}& \frac{1}{3}& \frac{2}{3}\\
\end{array}
        \right).
        \end{small}
\end{equation*}
The orthogonal classical quantum Latin squares of dimension 3 and 4 are
\begin{equation}
\begin{small}
\setlength{\arraycolsep}{5.0 pt}
\begin{array}{|c|c|c|c|}
\hline
|0\rangle & |1\rangle & |2\rangle  & |3\rangle \\
\hline
|3\rangle & |2\rangle & |1\rangle  & |0\rangle \\
\hline
|1\rangle & |0\rangle & |3\rangle  & |2\rangle \\
\hline
|2\rangle & |3\rangle & |0\rangle  & |1\rangle \\
\hline
\end{array}~~
\begin{array}{|c|c|c|c|}
\hline
|0\rangle & |1\rangle & |2\rangle  & |3\rangle \\
\hline
|2\rangle & |3\rangle & |0\rangle  & |1\rangle \\
\hline
|3\rangle & |2\rangle & |1\rangle  & |0\rangle \\
\hline
|1\rangle & |0\rangle & |3\rangle  & |2\rangle \\
\hline
\end{array}~~
\begin{array}{|c|c|c|}
\hline
|0\rangle & |1\rangle & |2\rangle \\
\hline
|1\rangle & |2\rangle & |0\rangle\\
\hline
|2\rangle & |0\rangle & |1\rangle \\
\hline
\end{array}~~
\begin{array}{|c|c|c|}
\hline
|0\rangle & |1\rangle & |2\rangle \\
\hline
|2\rangle & |0\rangle & |1\rangle\\
\hline
|1\rangle & |2\rangle & |0\rangle \\
\hline
\end{array}.
\end{small}\label{eq0}
\end{equation}
\hspace{2cm}$L^1$\hspace{3cm}$L^2$\hspace{2cm}$K^1$\hspace{2cm}$K^2$\\
Define $\Phi$ and $\Psi$ as the arrays (4) and (5), then $\Phi$ and $\Psi$ is a pair of orthogonal quantum Latin squares of dimension 12. Furthermore,  put $(i,j)=(0,3)$, $(m,n)=(9,10)$, then
 $|\langle\Phi_{0,3}|\Phi_{9,10}\rangle|=|\langle3|U|4\rangle|=|\frac{1-\sqrt{-1}}{\sqrt{6}}|\neq 0$ or~ $\neq 1$ for  $\Phi$;
put $(i,j)=(0,3)$, $(m,n)=(9,1)$, then $|\langle\Psi_{0,3}|\Psi_{9,1}\rangle|=|\langle 3|U^{\dagger}|4\rangle|=|\sqrt{\frac{-1}{6}}|\neq 0$ or $\neq$ 1 for $\Psi$, where we set $|3\rangle=|1\rangle\otimes|0\rangle$ and $|4\rangle=|1\rangle\otimes|1\rangle$. Thus $\Phi$ and $\Psi$ are both non-classical quantum Latin squares  by Lemma~\ref{classical}.
\begin{equation}
\begin{small}
\setlength{\arraycolsep}{2.0 pt}
\Phi=\begin{array}{|c|c|c|c|c|c|c|c|c|c|c|c|}
\hline
|0\rangle & |1\rangle & |2\rangle & |3\rangle & |4\rangle & |5\rangle &|6\rangle & |7\rangle & |8\rangle& |9\rangle & |10\rangle & |11\rangle  \\
\hline
|1\rangle & |2\rangle & |0\rangle &|4\rangle & |5\rangle & |3\rangle& |7\rangle & |8\rangle & |6\rangle& |10\rangle & |11\rangle & |9\rangle \\
\hline
|2\rangle & |0\rangle & |1\rangle & |5\rangle & |3\rangle & |4\rangle&|8\rangle & |6\rangle & |7\rangle&|11\rangle & |9\rangle & |10\rangle  \\
\hline
|9\rangle & |10\rangle & |11\rangle & |6\rangle & |7\rangle & |8\rangle & |3\rangle & |4\rangle & |5\rangle &  |0\rangle & |1\rangle & |2\rangle \\
\hline
|10\rangle & |11\rangle & |9\rangle& |7\rangle & |8\rangle & |6\rangle& |4\rangle & |5\rangle & |3\rangle&  |1\rangle & |2\rangle & |0\rangle \\
\hline
|11\rangle & |9\rangle & |10\rangle& |8\rangle & |6\rangle & |7\rangle& |5\rangle & |3\rangle & |4\rangle&  |2\rangle & |0\rangle & |1\rangle \\
\hline
|3\rangle & |4\rangle & |5\rangle & U|0\rangle & U|1\rangle & U|2\rangle  &|9\rangle & |10\rangle & |11\rangle & |6\rangle & |7\rangle & |8\rangle  \\
\hline
|4\rangle & |5\rangle & |3\rangle& U|1\rangle & U|2\rangle & U|0\rangle &|10\rangle & |11\rangle & |9\rangle &|7\rangle & |8\rangle & |6\rangle\\
\hline
|5\rangle & |3\rangle & |4\rangle& U|2\rangle & U|0\rangle & U|1\rangle &|11\rangle & |9\rangle & |10\rangle &|8\rangle & |6\rangle & |7\rangle \\
\hline
U|6\rangle & U|7\rangle & U|8\rangle &|9\rangle & |10\rangle & |11\rangle &|0\rangle & |1\rangle & |2\rangle & U|3\rangle & U|4\rangle & U|5\rangle \\
\hline
U|7\rangle & U|8\rangle & U|6\rangle& |10\rangle & |11\rangle & |9\rangle &|1\rangle & |2\rangle & |0\rangle &U|4\rangle & U|5\rangle & U|3\rangle \\
\hline
U|8\rangle & U|6\rangle & U|7\rangle&|11\rangle & |9\rangle & |10\rangle &|2\rangle & |0\rangle & |1\rangle & U|5\rangle & U|3\rangle & U|4\rangle \\
\hline
\end{array}
\end{small}
\end{equation}
\begin{equation}
\begin{small}
\setlength{\arraycolsep}{2.4 pt}
\Psi=\begin{array}{|c|c|c|c|c|c|c|c|c|c|c|c|}
\hline
|0\rangle & |1\rangle & |2\rangle & U|3\rangle & U|4\rangle & U|5\rangle &|6\rangle & |7\rangle & |8\rangle& |9\rangle & |10\rangle & |11\rangle  \\
\hline
|2\rangle & |0\rangle & |1\rangle &U|5\rangle & U|3\rangle & U|4\rangle& |8\rangle & |6\rangle & |7\rangle& |11\rangle & |9\rangle & |10\rangle \\
\hline
|1\rangle & |2\rangle & |0\rangle & U|4\rangle & U|5\rangle & U|3\rangle&|7\rangle & |8\rangle & |6\rangle&|10\rangle & |11\rangle & |9\rangle  \\
\hline
|6\rangle & |7\rangle & |8\rangle &|9\rangle & |10\rangle & |11\rangle &|0\rangle & |1\rangle & |2\rangle & |3\rangle & |4\rangle & |5\rangle \\
\hline
|8\rangle & |6\rangle & |7\rangle& |11\rangle & |9\rangle & |10\rangle &|2\rangle & |0\rangle & |1\rangle &|5\rangle & |3\rangle & |4\rangle \\
\hline
|7\rangle & |8\rangle & |6\rangle&|10\rangle & |11\rangle & |9\rangle &|1\rangle & |2\rangle & |0\rangle & |4\rangle & |5\rangle & |3\rangle \\
\hline
|9\rangle & |10\rangle & |11\rangle & U|6\rangle & U|7\rangle & U|8\rangle & |3\rangle & |4\rangle & |5\rangle &  U|0\rangle & U|1\rangle & U|2\rangle \\
\hline
|11\rangle & |9\rangle & |10\rangle& U|8\rangle & U|6\rangle & U|7\rangle& |5\rangle & |3\rangle & |4\rangle&  U|2\rangle & U|0\rangle & U|1\rangle \\
\hline
|10\rangle & |11\rangle & |9\rangle& U|7\rangle & U|8\rangle & U|6\rangle& |4\rangle & |5\rangle & |3\rangle&  U|1\rangle & U|2\rangle & U|0\rangle \\
\hline
|3\rangle & |4\rangle & |5\rangle & |0\rangle & |1\rangle & |2\rangle  &|9\rangle & |10\rangle & |11\rangle & |6\rangle & |7\rangle & |8\rangle  \\
\hline
|5\rangle & |3\rangle & |4\rangle& |2\rangle & |0\rangle & |1\rangle &|11\rangle & |9\rangle & |10\rangle &|8\rangle & |6\rangle & |7\rangle\\
\hline
|4\rangle & |5\rangle & |3\rangle& |1\rangle & |2\rangle & |0\rangle &|10\rangle & |11\rangle & |9\rangle &|7\rangle & |8\rangle & |6\rangle \\
\hline
\end{array}
\end{small}
\end{equation} \qed

From Lemma~\ref{md}, Lemma~\ref{Md0}, Corollary~\ref{mqod1d2} and Corollary~\ref{4.14}, we draw the following conclusion.


\begin{theorem}\label{Md1}

	1) Suppose that $d = p_{1}^{r_1}p_{2}^{r_2}\ldots p_{s}^{r_s} $, where $s\geq 2$, $r_i$ is a positive integer, $p_i$  is a prime and $p_i\neq p_j$ for $1\leq i\neq j\leq s$. Then
	$$M(d)\geq \min\{p^{r_i}_i-1 : 1\leq i\leq s \};$$
moreover, if $s=1,r_1\geq 2$, with $r_1=r'+r''$, $0< r'\leq r''$, then $M(d)\geq p^{r'}_1-1$.

2) Let $E_2 =\{2,3,4,6,8,18,p,2p,6p: p\geq 5 ~is~ a~ prime\}$. If $d\not \in E_2$, then $M(d)\geq 2$.

3) Let $E_3=\{9,12,24,27,50,54,3p: p\geq 5 ~is~ a~ prime\}$. If $d\not \in E_2 \cup E_3$, then $M(d)\geq 3$.

4) Let $E_4=\{16,32,36,48,66,110,242,4p:  p\geq 5 ~is~ a~ prime\}=E_4$. If $d\not \in E_2\cup E_3 \cup E_4$, then $M(d)\geq 4$.

\end{theorem}
\subsection{Filling-in-holes construction}

In this subsection, we will provide another method (filling in holes) to construct new types of orthogonal non-classical quantum Latin squares (self-orthogonal quantum Latin square), which are not equivalent to the orthogonal quantum Latin squares constructed in Section \ref{sec:directproduct}.
 Firstly, we review some basic combinatorial concepts
used in this work.

Let $H = \{S_1,S_2,\dots,S_n\}$ be a set of disjoint subsets of $ [d]$. An $incomplete~$ $Latin~$ $square$  ILS$(d; s_1, s_2,\dots , s_n)$~\cite{Colbourn} with hole set $H$ is a $d\times d$ array $L$ whose rows and columns are indexed by the elements of $ [d]$, and satisfies the following properties: 

1. each cell of $L$ is empty or contains an element of $ [d]$;

2. the subarrays (called holes) indexed by $S_i \times S_i$ are empty for $1\leq i \leq n$; and

3. suppose the row or column is indexed by $s$, then the elements in the row or column are exactly those of $[d]\setminus S_i$ if $s \in S_i$, and of $ [d]$ otherwise.

It is easy to see that when $H=\emptyset$, an incomplete~ Latin~ square is exactly a Latin~ square.

Two incomplete Latin squares on the symbol set $ [d]$ and with hole set $H$, say $L_1$ and $L_2$, are said to be $orthogonal$ and denoted by IMOLS$(d; s_1, s_2,\dots , s_n)$ if their superimposition yields every ordered pairs in $([d] \times [d])\setminus \bigcup_{i=1}^{n}(S_i \times S_i)$. Similarly, $t$-IMOLS$(d; s_1, s_2,\dots , s_n)$ denotes a set of $t$  ILS$(d; s_1, s_2,\dots , s_n)$s that are  pairwise orthogonal~\cite{Colbourn}.

If $H= \{S_1, S_2,\dots, S_n\}$ is a partition of $ [d]$, then an incomplete Latin square is called a $partitioned ~incomplete~ Latin~ square$, denoted by PILS. The $type$ of the PILS is defined to be the multiset $\{|S_i|:1\leq i\leq n\}$. We shall use an `exponential' notation to describe types, so type $h_1^{n_1}h_2^{n_2} \cdots h_l^{n_l}$ denotes $n_i$ occurrences of $h_i$, $1  \leq i \leq l$, in the multiset. Similarly, $t$-HMOLS$(h_1^{n_1}h_2^{n_2} \cdots h_l^{n_l})$ denotes a set of $t$ PILSs of type $h_1^{n_1}h_2^{n_2} \cdots h_l^{n_l}$ that are pairwise orthogonal~\cite{Bennett}.

An incomplete Latin square is called $self$-$orthogonal$ and denoted by ISOLS~\cite{Bennett}, if it is orthogonal to its transpose.
When $\{S_1, S_2,\dots, S_n\}$ is a partition of $ [d]$,  we use  the notation
HSOLS$(h_1^{n_1}h_2^{n_2} \cdots h_l^{n_l})$ instead of ISOLS with the type $h_1^{n_1}h_2^{n_2} \cdots h_l^{n_l}$ accurately.

\begin{lemma}\label{HSOLS}\cite{Stinson1,Stinson2}
For $h \geq 2$, there exists an HSOLS$(h^n)$ if and only if $n\geq 4$.
\end{lemma}
For more results on HMOLSs, we refer to~\cite{Bennett1,Bennett2,Dinitz,Lamken,Stinson,Stinson0}.

An incomplete Latin square is usually used to construct a Latin square by the method of filling in holes in the field of combinatorial designs. In this section, we are going to  construct some quantum Latin squares by applying that method. Now we generalize the definitions of ILS, $t$-IMOLSs, $t$-HMOLSs,  ISOLS and HSOLS to incomplete quantum Latin square, incomplete mutually orthogonal  quantum Latin squares and incomplete self-orthogonal  quantum Latin square.

\begin{definition}
 Let $V= \{V_1, V_2, \dots, V_n\}$ be a set of  mutually orthogonal subspaces of the complex vector space $\mathbb{C}^d$,  where $\dim V_i=d_i$ for $1\leq i \leq n$. An $incomplete~quantum~ Latin ~square$ IQLS$(d; d_1, d_2,\dots , d_n)$ with $hole~ set$ $V$ is a $d \times d$ array $\Psi$ whose rows and columns are indexed by one orthogonal basis of $\mathbb{C}^d$, $\{\phi_1,\phi_2,$ $\dots,\phi_d\}$,  satisfies the following properties:

1. every cell of $\Psi$ is either empty or contains a unit vector of $\mathbb{C}^d$;

2. the subarrays (called holes) whose rows and columns are indexed by the basis of $V_i$s  are empty; and

3. suppose the row or column is indexed by $\phi$, then the elements in the row or column are exactly the basis of  $\mathbb{C}^d \setminus V_i$ if $\phi \in V_i$, and of
$\mathbb{C}^d $ otherwise.
\end{definition}

An incomplete  classical Latin square is an incomplete quantum Latin square for which every element of the array is in the computational basis, and we call it a \emph{classical incomplete quantum Latin square}.

Two incomplete quantum Latin squares on $\mathbb{C}^d$  and hole set $V$, say $\Psi$ and $\Phi$, are said to be $orthogonal$, and are denoted by IMOQLS$(d; d_1, d_2,\dots, d_n)$,  if their ``superimposition" yields an orthonormal basis of $(\mathbb{C}^d \otimes \mathbb{C}^d )\setminus \bigoplus_{i=1}^{n}(V_i \otimes V_i)$. A set of $t$ IQLS$(d; d_1, d_2,\dots , d_n)$s that are  pairwise orthogonal is denoted by $t$-IMOQLS$(d; d_1, d_2,\dots, d_n)$.

Similar with the classical case, if $V= \{V_1, V_2,\dots, V_n\}$, where $\bigoplus_{1\leq i \leq n}V_i=\mathbb{C}^d$, then an incomplete~quantum~ Latin ~square is called a $partitioned ~incomplete~ quantum~\\ Latin~square$, and denoted by PIQLS.  A set of $t$ PIQLSs of type $d_1^{n_1}d_2^{n_2} \cdots d_l^{n_l}$ that are pairwise orthogonal is denoted by
$t$-HMOQLS$(d_1^{n_1}d_2^{n_2} \cdots d_l^{n_l})$. Here the meaning of the notation of type is analogous to the classical case. The type of a PIQLS is defined to be the multiset $\{\dim V_i: 1\leq i \leq n\}$.  So type $d_1^{n_1}d_2^{n_2} \cdots d_l^{n_l}$ denotes $n_i$ occurrences of $d_i$, $1  \leq i \leq l$, in the multiset.

Here we give an example of two quantum Latin squares obtained from two different kinds of incomplete quantum Latin squares by filling  the holes.
\begin{example}\label{QLS4} There is a non-classical QLS$(4)$ from a classical IQLS$(4;2)$, and a non-classical QLS$(7)$ from a classical PIQLS$(1^32^2)$.
\end{example}
$\centering {\begin{array}{c c}
\setlength{\arraycolsep}{2.8 pt}
~~~~\Phi=\begin{array}{lc}
\mbox{}&
\begin{array}{cccc}|0\rangle&|3\rangle&|1\rangle &|2\rangle  \end{array}\\
\begin{array}{c}|0\rangle \\|3\rangle\\ |1\rangle\\ |2\rangle\\ \end{array}&
\begin{array}{|c|c|c|c|}
\hline
            &           & |1\rangle& |2\rangle\\
\hline
           &             &|2\rangle& |1\rangle\\
\hline
|1\rangle & |2\rangle  & |0\rangle & |3\rangle  \\
\hline
|2\rangle & |1\rangle  &|3\rangle & |0\rangle \\
\hline
\end{array}
\end{array}
~~~~~~~~~~~~~~~\Phi'=\begin{array}{lc}
\mbox{}&
\begin{array}{cccc}|0\rangle~~~~&|3\rangle~~&|1\rangle  &|2\rangle  \end{array}\\
\begin{array}{c}|0\rangle \\|3\rangle\\ |1\rangle\\ |2\rangle\\ \end{array}&
\begin{array}{|c|c|c|c|}
\hline
 {\color{red} \frac{|0\rangle +|3\rangle}{\sqrt{2}}}&  {\color{red}\frac{|0\rangle -|3\rangle}{\sqrt{2}}}& |1\rangle& |2\rangle\\
\hline
 {\color{red} \frac{|0\rangle -|3\rangle}{\sqrt{2}}}&  {\color{red}\frac{|0\rangle +|3\rangle}{\sqrt{2}}}&  |2\rangle&|1\rangle \\
\hline
|1\rangle & |2\rangle  & |0\rangle & |3\rangle  \\
\hline
|2\rangle & |1\rangle  &|3\rangle & |0\rangle \\
\hline
\end{array}
\end{array}
\end{array}}$

$$\centering {\begin{array}{c c}
\setlength{\arraycolsep}{1.8 pt}

\hspace{-0.6cm}\Psi=\begin{array}{lc}
\mbox{}&
\begin{array}{ccccccc}|0\rangle &|1\rangle & |2\rangle&  |3\rangle&  |4\rangle & |5\rangle & |6\rangle  \end{array}\\
\begin{array}{ccccccc}|0\rangle \\|1\rangle\\ |2\rangle\\ |3\rangle\\ |4\rangle\\ |5\rangle \\ |6\rangle\end{array}&
\begin{array}{|c|c|c|c|c|c|c|c|}
\hline
            & |3\rangle & |4\rangle & |5\rangle & |6\rangle & |1\rangle & |2\rangle \\
\hline
|2\rangle &             & |5\rangle &|6\rangle& |0\rangle & |4\rangle & |3\rangle\\
\hline
|1\rangle & |6\rangle &            & |0\rangle & |5\rangle  & |3\rangle & |4\rangle\\
\hline
|6\rangle & |5\rangle & |1\rangle &            &            &|2\rangle & |0\rangle \\
\hline
|5\rangle & |2\rangle & |6\rangle &            &            &|0\rangle & |1\rangle \\
\hline
|3\rangle & |4\rangle & |0\rangle &|1\rangle & |2\rangle &            &            \\
\hline
|4\rangle & |0\rangle & |3\rangle &|2\rangle & |1\rangle &            &            \\
\hline
\end{array}
\end{array}
~~\Psi'=\begin{array}{lc}
\mbox{}&
\begin{array}{ccccccc}|0\rangle &|1\rangle & |2\rangle&~~~|3\rangle&~~~~|4\rangle &~~~|5\rangle &~~~|6\rangle  \end{array}\\
\begin{array}{ccccccc}|0\rangle \\|1\rangle\\ |2\rangle\\ |3\rangle\\ |4\rangle\\ |5\rangle \\ |6\rangle\end{array}&
\begin{array}{|c|c|c|c|c|c|c|c|}
\hline
{\color{red}  |0\rangle}    & |3\rangle & |4\rangle & |5\rangle & |6\rangle & |1\rangle & |2\rangle \\
\hline
|2\rangle &  {\color{red} |1\rangle}  & |5\rangle &|6\rangle& |0\rangle & |4\rangle & |3\rangle\\
\hline
|1\rangle & |6\rangle &   {\color{red} |2\rangle} & |0\rangle & |5\rangle  & |3\rangle & |4\rangle\\
\hline
|6\rangle & |5\rangle & |1\rangle &{\color{red} \frac{|3\rangle +|4\rangle}{\sqrt{2}}}&{\color{red} \frac{|3\rangle -|4\rangle}{\sqrt{2}}} &|2\rangle & |0\rangle \\
\hline
|5\rangle & |2\rangle & |6\rangle & {\color{red} \frac{|3\rangle -|4\rangle}{\sqrt{2}}}&{\color{red} \frac{|3\rangle +|4\rangle}{\sqrt{2}}}&|0\rangle & |1\rangle \\
\hline
|3\rangle & |4\rangle & |0\rangle &|1\rangle & |2\rangle &{\color{red} \frac{|5\rangle +|6\rangle}{\sqrt{2}}}& {\color{red} \frac{|5\rangle -|6\rangle}{\sqrt{2}}} \\
\hline
|4\rangle & |0\rangle & |3\rangle &|2\rangle & |1\rangle &{\color{red} \frac{|5\rangle-|6\rangle}{\sqrt{2}}}&{\color{red} \frac{|5\rangle +|6\rangle}{\sqrt{2}}}\\
\hline
\end{array}
\end{array}
\end{array}}$$

Notice that for a PIQLS we can always get diagonal holes by permuting the rows and columns, therewith the order of the indexes changed such as $\Phi$ or $\Psi$ in Example~\ref{QLS4}. Moreover, by the process of filling in holes in Example~\ref{QLS4}, the construction below can be obtained directly without proof.
\begin{construction}\label{PIQLSd1d2}(Filling in Holes)
If there exists an  IQLS$(d; d_1, d_2,\dots , d_n)$ and a QLS$(d_i)$ for $1\leq i\leq n$, then there exists a QLS$(d)$.
\end{construction}

\vspace{0.2cm}
An incomplete quantum Latin square is called $self$-$orthogonal$ if it is orthogonal to its conjugate transpose. We use the notation ISOQLS$(d; d_1, d_2,\dots, d_n)$ for  incomplete self-orthogonal quantum Latin square
and HSOQLS for when $\{V_1,V_2,\dots,\\V_n\}$ is a partition of $\mathbb{C}^d$ like classical ones.

From the Construction~\ref{PIQLSd1d2}, we  get the following corollary.
\begin{corollary}\label{QSd1d2}
If there exists an  HSOQLS$(d_1^{n})$ and a SOQLS$(d_1)$, then there exists a SOQLS$(d_1n)$.
\end{corollary}

In particular, we  can also obtain a non-classical  SOQLS($d_1n$) from a classical HSOQLS$(d_1^{n})$ by filling in the holes  of size $d_1$ with  SOQLS$(d_1)$s which are from  classical SOLS$(d_1)$s after a unitary matrix action. Here we denote by $\mathbb{Z}_d=\{0,1,\dots,d-1\}$  the additive group of integers modulo $d$.

\begin{construction}\label{Sd1d2}
If there exists an  HSOLS$(d_1^{n})$ and a SOLS$(d_1)$, then there exists an  SOQLS$(d_1n)$.
\end{construction}
See Appendix~\ref{sec:AppB} for the proof of Construction~\ref{Sd1d2}.

\begin{example}\label{Q16} (Non-classical SOQLS)
There exists a SOQLS$(16)$.
\end{example}

See Appendix~\ref{sec:AppI}  for the proof of Example~\ref{Q16}.

According to the constructions above,  a SOQLS can generate a pair of orthogonal quantum Latin squares. But it is easy to see that the SOQLS is not equivalent to a 2-MOQLSs constructed in Section~\ref{sec:directproduct},  since SOQLS has the special property that it is orthogonal with its transpose. Moreover, by Lemma~\ref{SOLS}, Lemma~\ref{HSOLS} and Construction~\ref{Sd1d2}, we  get the main result of this subsection.
\begin{theorem}\label{SQLSMd}
If $d_1,d_2\geq 4$, then there exists a SOQLS$(d_1d_2)$, except possibly for dimension 36.
\end{theorem}
\vspace{0.2cm}
Incomplete quantum Latin squares play an important role in the construction of filling in holes. Here we present a helpful construction for getting incomplete quantum Latin squares, which is a variation of the weighting construction of Lemma 3.6 in~\cite{Stinson1}.
\begin{construction}\label{HMOQLS} (Weighting)
If there exists a (classical) HMOQLS$(h^n)$ and a (non-classical) 2-MOQLS$(m)$, then there exists a (non-classical) HMOQLS$((hm)^n)$.
\end{construction}

See Appendix~\ref{sec:AppC} for the proof of Construction~\ref{HMOQLS}. Furthermore, Construction~\ref{HMOQLS} can be generalized to $t$-HMOQLSs.

\begin{corollary}\label{t-HMOQLS}
If there exists a (classical) $t$-HMOQLS$(h^n)$ and  a (non-classical) $t$-MOQLS$(m)$, then there exists a (non-classical) $t$-HMOQLS$((hm)^n)$.
\end{corollary}

Let $\Psi=\{|\Psi_{i,j}\rangle\}$ be an HSOQLS$(h^n)$ with hole set $V=\{V_1,V_2,\dots,V_n\}$ on $\mathbb{C}^{hn}$, and assume that the holes are in the diagonal line, and $\dim V_i=h$ for $1\leq i\leq n$. Suppose $\Phi^1=\{|\Phi^1_{l,k}\rangle\}$ and $\Phi^2=\{|\Phi^2_{l,k}\rangle\}$ is a 2-MOQLS$(m)$ on $\mathbb{C}^m$. Let $\Phi=\{|\Phi_{(i,l),(j,k)}\rangle\}$,
where
\begin{equation}
 |\Phi_{(i,l),(j,k)}\rangle= \left\{
          \begin{array}{ll}
|\Psi_{i,j}\rangle \otimes |\Phi^1_{l,k}\rangle,~ \mathrm{if}~ i\leq j;\\
|\Psi_{i,j}\rangle \otimes |\Phi^{2}_{k,l}\rangle^*, ~ \mathrm{otherwise}.\\
          \end{array}
       \right.\label{4.20}
\end{equation}
 Then $\Phi$ is an HSOQLS$((hm)^n)$ with hole set $V'=\{V_1\otimes \mathbb{C}^m, V_2\otimes \mathbb{C}^m,\dots, V_n\otimes\mathbb{C}^m\}$ on $\mathbb{C}^{hmn}$.

\begin{construction}\label{HSOQLS}
If there exists an HSOQLS$(h^n)$ and a  2-MOQLS$(m)$, then there exists an HSOQLS$((hm)^n)$.
\end{construction}
See Appendix~\ref{sec:AppD}  for the proof of Construction~\ref{HSOQLS}.

\begin{example}\label{3^4}
An HSOQLS$(3^4)$ can be constructed from an HSOQLS$(1^4)$ and a  2-MOQLS$(3)$.
\end{example}

See Appendix~\ref{sec:AppJ} for the proof of Example~\ref{3^4}.

\section{Quantum Latin cubes}
\label{sec:QLC}

\subsection{Classical Latin cubes}
In this section, we  list some notions of Latin cubes and the orthogonality among them which are from Ref.~\cite{Donald}.

A (\emph{classical $)$ Latin cube}  of order $d$, denoted by LC$(d)$, is a $d\times d \times d$ cube ($d$ rows, $d$ columns and $d$ files) in which the numbers $0,1,\dots, d - 1$ are entered so that each number occurs exactly once in each row, column and file. Three Latin cubes of order $d$ are \emph{orthogonal}, if when superimposed, each ordered triple $000, 001, \dots,d - 1\, d - 1\, d - 1$ occurs. A set of Latin cubes $L_1,L_2,\dots,L_t$ $(t\geq 3)$ is  mutually orthogonal, or a set of MOLC, if for every $1 \leq x < y< z\leq t$, $L_x$, $L_y$ and $L_z$ are orthogonal. We denote such set by $t$-MOLC($d$).

Mutually orthogonal classical Latin cubes have a close relationship with orthogonal arrays of strength 3 and $\lambda=1$.

\begin{lemma}\label{OA3}
For $t\geq 3$, there exists an $OA(d^3,t+3,d,3)$  if and only if after removing the first 3 columns,
 the remanning $t$ columns  satisfy the following  conditions:

$(A)$ they  correspond to $t$ mutually orthogonal Latin cubes;

$(B)$ every corresponding planes of any two cubes is a pair of orthogonal Latin squares.
\end{lemma}
See Appendix~\ref{sec:AppE}  for the proof of Lemma~\ref{OA3}. In the following,  we mainly consider the special case of $t$ mutually  orthogonal Latin cubes having Property (B).

Let  $c(d)$ be the largest number of mutually  orthogonal classical Latin cubes of order $d$ with Property (B). By the relation between MOLCs and OAs in Lemma~\ref{OA3}, some results about the number $c(d)$ follow.

\begin{lemma}\label{nd}(\cite{Donald,Colbourn,JiL})

	1) For any integer $d\geq 2$, $c(d)\leq d-1$.

	2) If $q\geq 5$ is a prime power, then $c(q)\geq q-2$. Moreover, if $q\geq 4$ is a power of $2$, then $c(q)\geq q-1$.

	3) Let $d$ be an integer  satisfying $\gcd(d,4)\neq 2$ and $\gcd(d,18) \neq 3$, then $c(d)\geq 3$.
Besides, $c(15),c(21)\geq 3$.

\end{lemma}

\subsection{Quantum Latin cubes}
Goyeneche et al.\ put forward the concepts of quantum Latin cube and orthogonality among three quantum Latin cubes in Ref.~\cite{Goyeneche1}. In this section, we review the concept and give a new definition of orthogonality among the quantum Latin cubes.

\begin{definition}\label{QLC}
A quantum Latin cube  $\Phi$ of dimension $d$, denoted by QLC$(d)$, is a $d\times d\times d$ cube  of elements $|\Phi_{i,j,k}\rangle \in \mathbb{C}^d , i,j,k \in [d]$, such that
every row, every column and every file determine an orthonormal basis of the complex Hilbert space $\mathbb{C}^d$.
\end{definition}

Two classical Latin cubes are said to be equivalent if one can be transformed into the other by
permutations of the rows, columns, files or relabeling of the symbols. Similarly, we give a notion of equivalence between two quantum Latin cubes.

\begin{definition}
Two quantum Latin cubes $\Phi$, $\Psi$ of dimension $d$ are equivalent if there exist a
unitary operator $U$ on $\mathbb{C}^d$, a family of modulus-1 complex numbers $c_{ijk}$, and three permutations $\sigma,\tau,\zeta \in S_d$, such
that the following holds for all $i,j,k \in [d]$:
\begin{equation}
|\Psi_{i,j,k}\rangle = c_{ijk}U|\Phi_{\sigma(i),\tau(j),\zeta(k)}\rangle.
\end{equation}
\end{definition}

A classical Latin cube can form a quantum Latin cube by associating each number in
 the classical Latin cube with a computational basis element, and we call it \emph{classical quantum Latin cube}. Moreover, if there is a quantum Latin cube equivalent to a classical one, then we also call it a classical quantum Latin cube, otherwise, it is a \emph{non-classical quantum Latin cube}. Similarly to classical quantum Latin squares,  classical quantum Latin cubes also have the following property, with a similar proof.
\begin{lemma}\label{classicalcube}
If~ $\Phi$ is a classical quantum Latin cube of dimension $d$, then for any~$i,j,k,f,g,h\in [d]$, one has ~$|\langle\Phi_{i,j,k}|\Phi_{f,g,h}\rangle|=0$ or~$1$.
\end{lemma}

Now we give a definition of mutually  orthogonal quantum Latin cubes, which differs from the one given in~\cite{Goyeneche1} by adding a condition similar to  property (B), and is analogous to Definition 11 of $m$ triplewise orthogonal quantum frequency cubes  in  Ref.~\cite{Pang}. This will establish a direct link of this notion  with that of a quantum orthogonal array in Definition~\ref{QOAdiyi}.
\begin{definition}\label{MOQLC1}
Three quantum Latin cubes $\Phi$,$\Psi$,$\Upsilon$ of dimension $d$ are orthogonal, if the following properties hold:

	1) $\{|\Phi_{i,j,k}\rangle\otimes |\Psi_{i,j,k}\rangle \otimes |\Upsilon_{i,j,k}\rangle: i,j,k\in [d ] \} $ forms an orthonormal basis of the space $\mathbb{C}^{d}\otimes \mathbb{C}^{d}\otimes \mathbb{C}^{d}$.

	2) for each fixed $i$,$j$ or $k$, the corresponding planes of any two cubes coming from $\Phi$,$\Psi$ and $\Upsilon$ can form a pair of orthogonal quantum Latin squares,
i.e.\  $$\sum_{xy}|\Lambda_{i,j,k} \rangle\langle\Lambda_{i,j,k}| \otimes |\Delta_{i,j,k}\rangle \langle\Delta_{i,j,k}|=\mathbb{I}_{d^2},$$ for different  $x,y\in \{i,j,k\}$, and $\Lambda,\Delta\in \{\Phi,\Psi,\Upsilon\}$.

\end{definition}

A set of $t\geq 3$ quantum Latin cubes of dimension $d$,  say $\Phi_1, \Phi_2,\dots, \Phi_t$,  is said to be mutually
orthogonal if $\Phi_i$, $\Phi_j$ and $\Phi_k$ are orthogonal for all $1 \leq i < j<k \leq t$, and is denoted by $t$-MOQLC$(d)$.

Thus a set of mutually  orthogonal classical Latin cubes with Property (B)  forms a set of mutually orthogonal classical quantum Latin cubes.

Let $C(d)$ be the largest number of mutually orthogonal non-classical quantum Latin cubes of dimension $d$. From Lemma~\ref{Md0} and  condition (2) in Definition~\ref{MOQLC1}, we can establish the following  upper bound.
\begin{lemma}\label{Cd0}
For any $d\geq 2$, $C(d)\leq d-1$.
\end{lemma}

\begin{definition}
 Given a quantum Latin cube $\Phi$, its conjugate $\Phi^*$, is the quantum Latin cube with entries
$(|\Phi^*_{i,j,k}\rangle) = (|\Phi_{i,j,k}\rangle^*)$.
\end{definition}
In~\cite{Musto2}, it is proved the orthogonality of quantum Latin squares is unaffected by conjugation of one of the squares. For quantum Latin cubes, an analogous result holds.
\begin{lemma}\label{psi*}
Three quantum Latin cubes $\Phi$, $\Psi$ and $\Upsilon$ are orthogonal, if and only if $\Phi^*$, $\Psi$ and $\Upsilon$ or  $\Phi^*$, $\Psi^*$ and $\Upsilon$  are orthogonal.
\end{lemma}
See Appendix~\ref{sec:AppF}  for the proof of Lemma~\ref{psi*}.

\subsection{Direct product construction}

In this subsection, we will provide a direct product construction of mutually  orthogonal quantum Latin cubes. In particular, we will give a method to construct  mutually  orthogonal non-classical quantum Latin cubes from  mutually  orthogonal  classical Latin cubes with Property (B).

The following direct product construction is similar to the construction described in Section 2.3, and we will not give the proof  here.

\begin{construction}\label{qocd1d2}(Direct Product Construction)
If there exists a 3-$MOQLC(d_1)$ and a 3-$MOQLC(d_2)$, then there exists a 3-$MOQLC(d_1d_2)$.
\end{construction}

\begin{corollary}\label{MOQLCd1d2}
Let $l\geq 2$. If there exists a $t_j$-MOQLC$(d_j)$, for any $1\leq j\leq l$,  then there exists  a $t$-MOQLC$(d)$, where $t= \min\{t_1,t_2,\dots,t_l\}$ and $d=d_1d_2\cdots d_l$.
\end{corollary}

\begin{construction}\label{cd1d2}
If there exists a  3-$MOLC(d_1)$ and a  3-$MOLC(d_2)$ both with Property (B), then  there exists a 3-$MOQLC(d_1d_2)$.
\end{construction}

See Appendix~\ref{sec:AppG}  for the proof of Construction~\ref{cd1d2}. Obviously, from the proof we cannot choose $\tau$s all being $I$ or $U$, if we want to get a non-classical quantum Latin cubes.

\begin{corollary}\label{0MOQLCd1d2}
Let $l\geq 2$ and $d=d_1d_2\cdots d_{l}$, with $ c(d_j)\geq 3$ for all $1\leq j\leq l$. Then there exists a $t$-MOQLC$(d)$ with $t=\min\{c(d_1),c(d_2),\dots, c(d_l)\}$.
\end{corollary}

\begin{example}\label{moqlcs16}(Non-classical MOQLCs)
There exists a 3-MOQLC(16).
\end{example}

See Appendix~\ref{sec:AppK} for the proof of Example~\ref{moqlcs16}.

By Lemma~\ref{nd}, Corollary~\ref{MOQLCd1d2}  and Corollary~\ref{0MOQLCd1d2}, we finally get the following theorem.
\begin{theorem}\label{Cd1}
	1) Suppose that $d = p_{1}^{r_1}p_{2}^{r_2}\ldots p_{s}^{r_s} $, where $s\geq 2$, $r_i$ is a positive integer, $p_i$  is a prime and $p_i\neq p_j$ for $1\leq i\neq j\leq s$, then $C(d)\geq \min\{p^{r_i}_i-2 : 1\leq i\leq s\}$; if $s=1$, $r_1\geq 2$, with $r_1=r'+r''$, $0<r'\leq r''$, then $C(d)\geq p^{r'}_1-2$.

	2)  Let $d_1$, $d_2$ be integers  satisfying $\gcd(d_i,4)\neq 2$ and $\gcd(d_i,18) \neq 3$,  $i\in \{1,2\}$, then $C(d_1d_2)\geq 3$.

\end{theorem}

\section{Generalized orthogonality for QLSs and QLCs}
\label{sec:GQLS}

In 2018, Goyeneche et al. put forward the notion of quantum orthogonal array~\cite{Goyeneche1}, which allows to  obtain a $k$-uniform state from a QOA.  Moreover, they point out the close relation between QOAs and mutually  orthogonal quantum Latin squares (or cubes). In this section, we  elaborate on the notions in~\cite{Goyeneche1} of orthogonality among quantum Latin squares (cubes) whose arrangements may be entangled.  Furthermore, we show a one-to-one relationship between them and QOAs  of  strength $2, 3$ with minimal support, such that a family of $k$-uniform states for $k=2, 3$ can be derived by combining the results of the previous sections. The definition of QOA here is the same as that of IQOA given in Ref.~\cite{Du}.

Let $(\mathbb{C}^d)^{\otimes N}=\mathbb{C}^d \otimes \mathbb{C}^d  \cdots \otimes \mathbb{C}^d $, be the $N$-fold tensor product of $\mathbb{C}^d$.
The unit vectors belonging to $(\mathbb{C}^d)^{\otimes N}$ represents pure quantum states of $N$ parties having $d$ internal levels each.
\begin{definition}\label{QOAdiyi}
 A quantum orthogonal array $QOA(r,N,d,k)$ is an arrangement consisting of $r$ rows composed by $N$-partite  pure quantum states $|\varphi_{i}\rangle\in (\mathbb{C}^d)^{\otimes N}$
 such that,
\begin{eqnarray}
\sum_{i,j=0}^{r-1}Tr_{l_{1},\ldots,l_{N-k}}(|\varphi_{i}\rangle \langle \varphi_{j}|)
&=\frac{r}{d^{k}} \mathbb{I}_{d^{k}}.
\end{eqnarray}
for every subset $\{l_{1},\ldots,l_{N-k}\}$ of $N-k$ parties.
\end{definition}

Now we give a notion of generalized  orthogonality for QLSs and QLCs which allows to establish their equivalence to QOAs of strength 2, 3 with minimal support, i.e.\ $r=d^k$.

\begin{definition}\label{MOQLS2}
Let $t\geq 2$. A set of $d^2$ $t$-partite pure quantum states $|\psi_{i,j}\rangle \in (\mathbb{C}^d)^{\otimes t}$ arranged as
\begin{equation*}
\begin{array}{ccc}
|\psi_{0,0}\rangle &\cdots&|\psi_{0,d-1}\rangle   \\
\vdots & &\vdots   \\
|\psi_{d-1,0}\rangle &\cdots &|\psi_{d-1,d-1}\rangle \\
\end{array}
\end{equation*}
forms a set of \emph{generalized mutually orthogonal quantum Latin squares of dimension $d$}, denoted by $t$-GMOQLS$(d)$, 
if the following properties hold:

1. the $d^2$ states $|\psi_{i,j}\rangle$ are orthogonal, i.e.\
\begin{equation}
\langle\psi_{i,j}|\psi_{i',j'}\rangle=\delta_{ii'}\delta_{jj'}.  \label{eq1}
\end{equation}
\quad  2.
	\begin{equation}
\sum^{d-1}_{i=0}Tr_{l_1,l_2,\dots,l_{t-1}}|\psi_{i,j}\rangle\langle\psi_{i,j'}|=\delta_{jj'}\mathbb{I}_d, \label{eq2}
\end{equation}
\begin{equation}
\sum^{d-1}_{j=0}Tr_{l_1,l_2,\dots, l_{t-1}}|\psi_{i,j}\rangle\langle\psi_{i',j}|=\delta_{ii'}\mathbb{I}_d, \label{eq3}
\end{equation}
for every subset  $\{l_1,l_2,\dots ,l_{t-1}\}$ of $t-1$ parties.

 3.
	\begin{equation}
\sum^{d-1}_{i,j=0}Tr_{l_1,l_2,\dots, l_{t-2}}|\psi_{i,j}\rangle\langle\psi_{i,j}|=\mathbb{I}_{d^2}, \label{eq4}
\end{equation}
 for every subset  $\{l_1,l_2,\dots ,l_{t-2}\}$ of $t-2$ parties.
\end{definition}

\begin{remark}
	If a $t$-GMOQLS$(d)$ is composed of fully separable states, i.e.\ $|\psi^{A_1A_2\cdots A_t}_{i,j}\rangle\\=|\psi^{A_1}_{i,j}\rangle\otimes|\psi^{A_2}_{i,j}\rangle\otimes\cdots \otimes |\psi^{A_t}_{i,j}\rangle$ for every $i,j \in [d]$, then the $t$-GMOQLS$(d)$ is just a $t$-MOQLS$(d)$. In fact, Property (3) implies Property (1); Property (2) is equivalent with arrangement $\{|\psi^{A_s}_{i,j}\rangle\}$ being a QLS for every $s, 1 \leq s \leq t$, according to Definition~\ref{QLS}; and Property (2) and (3) are equivalent with  arrangements  $\{|\psi^{A_1}_{i,j}\rangle\},\{|\psi^{A_2}_{i,j}\rangle\},\dots, \{|\psi^{A_t}_{i,j}\rangle\}$ being  a set of $t$-MOQLS($d$) according to Definition~\ref{MOQLS1}.
\end{remark}

\begin{proposition}\label{GMOQLS}
A QOA$(d^2,t+2,d,2)$ generates a $t$-GMOQLS$(d)$, and vice versa.
\end{proposition}
\noindent \p  Suppose that $|\Phi\rangle$ is the sum of the $d^2$ states in  the QOA$(d^2,t+2,d,2)$. Since $|\Phi\rangle$ can  produce a 2-uniform state, we choose the first two subsystems, namely, $i$, $j$, then
\begin{equation}
|\Phi\rangle=\sum^{d-1}_{i,j=0}|ij\rangle\otimes|\Phi_{i,j}\rangle.
\end{equation}

Actually, $\{|\Phi_{i,j}\rangle\}$ is a set of arrangements of $t$-GMOQLS$(d)$, where $i,j\in [d]$ are the indexes of the rows and columns of the $t$-GMOQLS.
To show this, we take an arbitrary subset $S \subseteq \{1,2,\dots,t+2\}$ with $|S|=2$.  Consider the following three cases: (1) $|S\cap\{1,2\}|=2$; (2) $|S\cap \{1,2\}|=1$; (3) $|S\cap \{1,2\}|=0$.
\begin{itemize}
	\item[1.]
When $|S\cap\{1,2\}|=2$, we have
\begin{eqnarray*}
\rho_{S}&=Tr_{3,4,\dots,t+2}\sum_{i,j,i',j'\in [d]}|ij\rangle\otimes|\Phi_{i,j}\rangle \langle i'j'|\otimes\langle\Phi_{i',j'}|\\
&=\sum_{i,j,i',j'\in [d]}|ij\rangle\langle i'j'| \langle\Phi_{i',j'}|\Phi_{i,j}\rangle.
\end{eqnarray*}
 Thus, $\rho_{S}=\sum\limits_{i,j,i',j'\in [d]}|ij\rangle\langle i'j'| \langle\Phi_{i',j'}|\Phi_{i,j}\rangle= \mathbb{I}_{d^{2}}$ if and only if  Eq.~(\ref{eq1}) holds.

\item[2.]
When $|S\cap \{1,2\}|=1$. Suppose $S\cap \{1,2\}=\{1\}$, and $\{l_1,l_2,\dots, l_{t-1}\}\cap \{1,2\}=\emptyset$, then we have
\begin{eqnarray*}
\rho_{S}&=Tr_{2,l_1,l_2,\dots, l_{t-1}}\sum_{i,j,i',j'\in [d]}|ij\rangle\otimes|\Phi_{i,j}\rangle \langle i'j'|\otimes\langle\Phi_{i',j'}|\\
&=\sum_{i,i'\in [d]}|i\rangle\langle i'| \otimes \sum_{j\in [d]} Tr_{l_1,l_2,\dots, l_{t-1}} |\Phi_{i,j}\rangle \langle\Phi_{i',j}|.
\end{eqnarray*}
 So, $\rho_{S}=\sum\limits_{i,i'\in [d]}|i\rangle\langle i'| \otimes \sum\limits_{j\in [d]} Tr_{l_1,l_2,\dots, l_{t-1}} |\Phi_{i,j}\rangle \langle\Phi_{i',j}|= \mathbb{I}_{d^{2}}$ if and only if  Eq.~(\ref{eq3}) holds.
One the other hand, suppose $S\cap \{1,2\}=\{2\}$, then  $\rho_{S}=\sum\limits_{j,j'\in [d]}|j\rangle\langle j'| \otimes \sum\limits_{i\in [d]} Tr_{l_1,l_2,\dots, l_{t-1}} |\Phi_{i,j}\rangle \langle\Phi_{i,j'}|= \mathbb{I}_{d^{2}}$ if and only if  Eq.~(\ref{eq2}) holds.

\item[3.]
When $|S\cap \{1,2\}|=0$, we have
\begin{eqnarray*}
\rho_{S}&=Tr_{1,2,l_1,l_2,\dots, l_{t-2}}\sum_{i,j,i',j'\in [d]}|ij\rangle\otimes|\Phi_{i,j}\rangle \langle i'j'|\otimes\langle\Phi_{i',j'}|\\
&=\sum_{i,j\in [d]} Tr_{l_1,l_2,\dots, l_{t-2}} |\Phi_{i,j}\rangle \langle\Phi_{i,j}|.
\end{eqnarray*}
Thus, $\rho_{S}= \mathbb{I}_{d^{2}}$ if and only if  Eq.~(\ref{eq4}) holds. \qed
\end{itemize}

\begin{example}
	Consider the following quantum orthogonal array consisting of five columns~\cite{Goyeneche1}:
\begin{equation}
QOA(4,3_C+2_Q,2,2) =\left(
\begin{array}{cccc}
|0\rangle & |0\rangle & |0\rangle &|\Phi^+\rangle\\
|0\rangle & |1\rangle & |1\rangle &|\Psi^+\rangle\\
|1\rangle & |0\rangle & |1\rangle &|\Psi^-\rangle\\
|1\rangle & |1\rangle & |0\rangle &|\Phi^-\rangle\\
\end{array}
\right),
\end{equation}
where $|\Phi^{\pm}\rangle = (|00\rangle \pm |11\rangle)/\sqrt{2}$ and $|\Psi^{\pm}\rangle = (|01\rangle \pm |10\rangle)/\sqrt{2}$ are the Bell basis. We can see that the first three columns are separable (classical) and the last two columns are entangled (quantum). From  Proposition~\ref{GMOQLS}, let the first and second columns be the address of a triple of generalized mutually orthogonal quantum Latin squares. Then we  get
\begin{equation}
3-GMOQLS(2) =
\begin{array}{|c|c|}
\hline
|0\rangle |\Phi^+\rangle & |1\rangle |\Psi^+\rangle\\
\hline
|1\rangle |\Psi^-\rangle & |0\rangle |\Phi^-\rangle\\
\hline
\end{array}.
\end{equation}
\end{example}

\begin{definition}\label{MOQLC2}
Let $t\geq 3$. A set of $d^3$ $t$-partite pure quantum states $|\psi_{i,j,k}\rangle \in (\mathbb{C}^d)^{\otimes t}$ arranged as
$$\includegraphics[width=9cm,height=6cm]{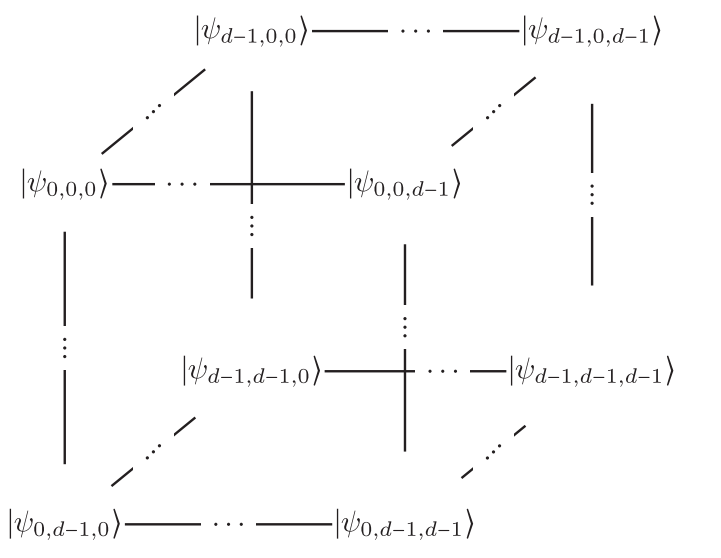}$$
forms a set of \emph{generalized mutually orthogonal quantum Latin cubes of dimension $d$}, denoted by $t$-GMOQLC$(d)$, if the following properties hold:

	1.
	the $d^3$ states $\{|\psi_{i,j,k}\rangle\}$ are orthogonal, i.e.\
\begin{equation}
\langle\psi_{i,j,k}|\psi_{i',j',k'}\rangle=\delta_{ii'}\delta_{jj'}\delta_{kk'}.  \label{eq1'}
\end{equation}

	2.
 \begin{equation}
\sum^{d-1}_{i=0}Tr_{l_1,l_2,\dots, l_{t-1}}|\psi_{i,j,k}\rangle\langle\psi_{i,j',k'}|=\delta_{jj'}\delta_{kk'}\mathbb{I}_d, \label{eq2'}
\end{equation}
\begin{equation}
\sum^{d-1}_{j=0}Tr_{l_1,l_2,\dots ,l_{t-1}}|\psi_{i,j,k}\rangle\langle\psi_{i',j,k'}|=\delta_{ii'}\delta_{kk'}\mathbb{I}_d, \label{eq3'}
\end{equation}
\begin{equation}
\sum^{d-1}_{k=0}Tr_{l_1,l_2,\dots ,l_{t-1}}|\psi_{i,j,k}\rangle\langle\psi_{i',j',k}|=\delta_{ii'}\delta_{jj'}\mathbb{I}_d, \label{eq4'}
\end{equation}
 for every subset $\{l_1,l_2,\dots ,l_{t-1}\}$ of $t-1$ parties.

	3.
 \begin{equation}
\sum^{d-1}_{i,j=0}Tr_{l_1,l_2,\dots,l_{t-2}}|\psi_{i,j,k}\rangle\langle\psi_{i,j,k'}|=\delta_{kk'}\mathbb{I}_{d^2}, \label{eq5'}
\end{equation}
\begin{equation}
\sum^{d-1}_{j,k=0}Tr_{l_1,l_2,\dots,l_{t-2}}|\psi_{i,j,k}\rangle\langle\psi_{i',j,k}|=\delta_{ii'}\mathbb{I}_{d^2}, \label{eq6'}
\end{equation}
\begin{equation}
\sum^{d-1}_{i,k=0}Tr_{l_1,l_2,\dots,l_{t-2}}|\psi_{i,j,k}\rangle\langle\psi_{i,j',k}|=\delta_{jj'}\mathbb{I}_{d^2}, \label{eq7'}
\end{equation}
 for every subset $\{l_1,l_2,\dots,l_{t-2}\}$ of $t-2$ parties.

	4.
	\begin{equation}
\sum^{d-1}_{i,j,k=0}Tr_{l_1,l_2,\dots,l_{t-3}}|\psi_{i,j,k}\rangle\langle\psi_{i,j,k}|=\mathbb{I}_{d^3}. \label{eq8'}
\end{equation}
 for every subset  $\{l_1,l_2,\dots,l_{t-3}\}$ of $t-3$ parties.
\end{definition}

\begin{remark}
	If a $t$-GMOQLC$(d)$ is composed of fully separable states, i.e.\ $|\psi^{A_1A_2\cdots A_t}_{i,j,k}\rangle\\=|\psi^{A_1}_{i,j,k}\rangle\otimes|\psi^{A_2}_{i,j,k}\rangle\otimes\cdots \otimes |\psi^{A_t}_{i,j,k}\rangle$ for every $i,j,k\in [d]$, then the $t$-GMOQLC$(d)$ is just a $t$-MOQLC$(d)$.  In fact, Property (4) implies Property (1); Property (2) implies that the arrangement $\{|\psi^{A_s}_{i,j,k}\rangle\}$ is a QLC, for every $s, 1 \leq s \leq t$, according to Definition~\ref{QLC}; Property (3) implies that for each corresponding planes of any two cubes from $|\psi^{A_1}_{i,j,k}\rangle,|\psi^{A_2}_{i,j,k}\rangle,\dots, |\psi^{A_t}_{i,j,k}\rangle$  form a pair of orthogonal quantum Latin square, which is consistent with Property (2) of Definition~\ref{MOQLC1}; and Property (4) implies that the ``superimposed" elements of any three cubes  form an orthonormal basis of $(\mathbb{C}^d)^{\otimes 3}$,  which is consistent with Property (1) of Definition~\ref{MOQLC1}.
\end{remark}

\begin{proposition}\label{GMOQLC}
A QOA$(d^3,t+3,d,3)$ generates a $t$-GMOQLC$(d)$, and vice versa.
\end{proposition}

See Appendix~\ref{sec:AppH} for the proof of Proposition~\ref{GMOQLC}. Here we give an example to show the relation between QOA and GMOQLC.

\begin{example}
A 4-GMOQLC(7) can be obtained from a QOA$(343,4_C+3_Q,7,3)$.
\end{example}
\noindent \p Let
\begin{equation}
QOA(343,4_C+3_Q,7,3) =\left(
\begin{array}{c}
|\Phi_{0,0,0}\rangle\\
|\Phi_{0,0,1}\rangle\\
\vdots\\
|\Phi_{6,6,6}\rangle\\
\end{array}
\right),
\end{equation}
where $ |\Phi_{i,j,k}\rangle=|i,k,i+j+k,i+2j+4k\rangle \otimes |\phi_{i,j,k}\rangle, ~|\phi_{i,j,k}\rangle=\frac{1}{\sqrt{7}}\sum\limits_{l=0}^{6}\omega^{il}|l+j,l+2j+5k,l\rangle,~0\leq i,j,k\leq 6$, and $\omega=e^{\frac{2\pi\sqrt{-1}}{7}}$. In the same way as in Proposition~\ref{GMOQLC}, let the first three columns be the address of a 4-tuple of generalized mutually orthogonal quantum Latin cubes, then we  get a 4-GMOQLC$(7)$ from the last four systems of the QOA$(343,4_C+3_Q,7,3)$.
\qed

\vspace{0.2cm}
So far, plenty of 2- and 3-uniform states have been obtained such as  2-uniform states for any $d\geq 2$, $N\geq 4$, except for $d=2, N=4$ \cite{Scott,Goyeneche0,Li,Zang1,Higuchi,Pang1,Rather,Rains};  3-uniform states for any $d\geq 2$,$N\geq 6$ except for $d\equiv 2$ (mod 4), $N=7$ \cite{Helwig,Huber1,Li,Zang,Pang1,Rains}; especially, AME(4,$d$) for $d\neq 2$, AME(5,$d$) for any $d$, AME(6,$d$) for any $d$ and  AME(7,$d$) for $d\not \equiv 2$ (mod 4) \cite{Huber1,Huber,Scott,Pang1,Li,Rains}. Strikingly, Pang et al. constructed the 2- and 3-uniform states for almost any $N$ and $d$, especially on AME($N$,$d$) for $N=4,5,6$,  by a special kind of orthogonal arrays \cite{Pang1}. From Propositions~\ref{GMOQLS} and~\ref{GMOQLC}, we  see that the MOQLSs and MOQLCs defined in Sections~\ref{sec:QLS} and~\ref{sec:QLC}  also have equivalent relations with  QOAs when they have  columns of fully separable states, with $k=2,3$ respectively, exactly like the classical ones in Lemmas~\ref{OA2} and~\ref{OA3}. Therefore, from  these relations we immediately obtain a method for constructing 2- and 3-uniform states with minimal-support which are not locally equivalent to the ones obtained from classical orthogonal arrays in~\cite{Pang1}.

\section{Conclusions}
\label{sec:concl}

A generalization of classical combinatorial arrangements to quantum information has been established. Musto and Vicary gave the notions of quantum Latin squares and the orthogonality on them~\cite{Musto1,Musto3}. Then Goyeneche et al. put forward the concepts of quantum Latin cubes and the orthogonality on them~\cite{Goyeneche1}.

In this article, we elaborated on the notion of mutually  orthogonal quantum Latin cubes. Since the arrangements of  MOQLSs and MOQLCs may be entangled, we came up with the notions of generalized mutually orthogonal quantum Latin squares and generalized mutually orthogonal quantum Latin cubes. In particular, MOQLSs and MOQLCs are  extreme cases of them with columns of fully separable states. Furthermore, we established one-to-one relationships between those GMOQLSs and QOAs, as well as GMOQLCs and QOAs. Meanwhile, we provided explicit construction methods of MOQLSs and MOQLCs by direct product and by filling in holes,  which in turn produce multipartite entangled $k$-uniform states for $k=2,3$.

A necessary condition for the existence of $k$-uniform states is $k\leq \lfloor N/2\rfloor$. From  Theorems~\ref{Md1},~\ref{SQLSMd} and Theorem~\ref{Cd1}, we  get new information on the properties of multipartite entanglement in $k$-uniform states and in particular on AME states. These are given by the following three theorems which represent
 the main conclusions of this work.

\begin{theorem} \label{thm:5.1}
	1) Suppose that $d = p_{1}^{r_1}p_{2}^{r_2}\ldots p_{s}^{r_s} $, where $s\geq 2$, $r_i$ is a positive integer, $p_i$  is a prime such that $p_{i}^{r_i}\geq 3$ for all $1\leq i\leq s$, and $p_i\neq p_j$ for $1\leq i\neq j\leq s$.
	Then there exists a 2-uniform state of $\min\{p^{r_i}_i+1 : 1\leq i\leq s \}$ subsystems with dimension $d$;
	
moreover, if $s=1,r_1\geq 2$, with $r_1=r'+r''$, $0< r'\leq r''$,  then there is a 2-uniform state of $p^{r'}_1+1$ subsystems.
2) If $d\not \in E_2$,  there is an AME$(4,d)$.

3) If $d\not \in E_2 \cup E_3$, there is an AME$(5,d)$.

4) If $d\not \in E_2 \cup E_3\cup E_4$, then there is a 2-uniform state of $6$ subsystems with dimension $d$.\\

\vspace{-0.4cm}\noindent Here $E_2, E_3$ and $E_4$ are the sets defined in Theorem~\ref{Md1}.
\end{theorem}

\begin{theorem}
If $d_1,d_2\geq 4$, then there is an AME$(4,d_1d_2)$, different from the one in Theorem~\ref{thm:5.1}, except possibly for dimension 36.
\end{theorem}

\begin{theorem}
	1) Suppose that $d = p_{1}^{r_1}p_{2}^{r_2}\ldots p_{s}^{r_s} $, where $s\geq 2$, $r_i$ is a positive integer, $p_i$  is a prime such that $p_{i}^{r_i}\geq 5$, for all $1\leq i\leq s$, and $p_i\neq p_j$ for $1\leq i\neq j\leq s$.
	 Then there is a 3-uniform state of $\min\{p^{r_i}_i+1 : 1\leq i\leq s\}$ subsystems with dimension $d$;
	
	 moreover, if $s=1$, $r_1\geq 2$, with $r_1=r'+r''$, $0<r'\leq r''$, then there is a 3-uniform state of $p^{r'}_1+1$ subsystems with dimension $d$.
	2)  Let $d_1$, $d_2$ be integers  satisfying $\gcd(d_i,4)\neq 2$ and $\gcd(d_i,18) \neq 3$,  $i\in \{1,2\}$. Then  there is an AME$(6,d_1 d_2)$.
\end{theorem}

 In this article, we have given explicit construction methods of  2- and 3-uniform states from MOQLSs and MOQLCs which can also be used to construct unitary error bases and mutually unbiased bases.  Recently, Peng constructed $k$-uniform states starting from QOAs~\cite{Peng} whose rows consist of entangled states, which are different from the ones exhibited here.  As a matter of fact, as shown in this work, establishing  alternative construction methods of GMOQLSs, GMOQLCs and QOAs has interesting and immediate applications in entanglement theory and in quantum information science.

\section*{Acknowledgments}
This work was supported by the National Natural Science Foundation of China under Grant No.\ 11871019 (Z. Tian). P.F. was partially supported by Istituto Nazionale di Fisica Nucleare (INFN) through the project ``QUANTUM'' and  by the Italian National Group of Mathematical Physics (GNFM-INdAM). Y. Zang  acknowledges the hospitality of the PhD school in Physics at the Physics Department of the University of Bari. This work was developed under the scientific cooperation and exchange program of Universit\`{a} degli Studi di Bari and Hebei Normal University. The authors thank both referees for their constructive suggestions and comments that allow for
notable improvements to the manuscript.

\appendix
\section{Proof of Construction~\ref{d1d2}}
\label{sec:AppA}

\noindent \p  Suppose $\mathbb{C}^{d_1}=\Span \{|0\rangle,|1\rangle,\dots,|d_1-1\rangle\}$ and $\mathbb{C}^{d_2}=\Span\{|0\rangle,|1\rangle,\dots,|d_2-1\rangle\}$. Then $\mathbb{C}^{d_1d_2}\simeq \mathbb{C}^{d_1}\otimes\mathbb{C}^{d_2}=\Span\{|i\rangle \otimes |j\rangle: i\in [d_1], j\in [d_2]
\}=\Span\{|0\rangle,|1\rangle,\dots,\\|d_1d_2-1\rangle\}$.

Let $l^1=(l^1_{i,j})_{d_1\times d_1}$, $l^2=(l^2_{i,j})_{d_1\times d_1}$ be a 2-MOLS$(d_1)$ and $k^1=(k^1_{m,n})_{d_2\times d_2}$, $k^2=(k^2_{m,n})_{d_2\times d_2}$ be a 2-MOLS$(d_2)$. Then put $L^1=\{|l^1_{i,j}\rangle: i,j\in [d_1]\}$, $L^2=\{|l^2_{i,j}\rangle: i,j\in [d_1]\}$, $K^1=\{|k^1_{m,n}\rangle: m,n\in [d_2]\}$, and $K^2=\{|k^2_{m,n}\rangle: m,n\in [d_2]\}$ to be the corresponding classical quantum Latin squares as follows.

\begin{equation}
\begin{small}
\setlength{\arraycolsep}{3.0pt}
L^1=\begin{array}{|c|c|c|}
\hline
|l^1_{0,0}\rangle  & \cdots & |l^1_{0,d_{1}-1}\rangle  \\
\hline
\cdots  &\cdots & \cdots  \\
\hline
|l^1_{d_{1}-1,0}\rangle  &\cdots& |l^1_{d_{1}-1,d_{1}-1}\rangle  \\
\hline
\end{array}
~~~~~~~~
L^2=\begin{array}{|c|c|c|}
\hline
|l^2_{0,0}\rangle  & \cdots & |l^2_{0,d_{1}-1}\rangle  \\
\hline
\cdots  &\cdots & \cdots  \\
\hline
|l^2_{d_{1}-1,0}\rangle  &\cdots& |l^2_{d_{1}-1,d_{1}-1}\rangle  \\
\hline
\end{array}
\end{small}
\end{equation}
\begin{equation}
\begin{small}
\setlength{\arraycolsep}{3.0pt}
K^1=\begin{array}{|c|c|c|}
\hline
|k^1_{0,0}\rangle  & \cdots & |k^1_{0,d_{2}-1}\rangle  \\
\hline
\cdots  &\cdots & \cdots  \\
\hline
|k^1_{d_{2}-1,0}\rangle  &\cdots& |k^1_{d_{2}-1,d_{2}-1}\rangle  \\
\hline
\end{array}
~~~~~
K^2=\begin{array}{|c|c|c|}
\hline
|k^2_{0,0}\rangle  & \cdots & |k^2_{0,d_{2}-1}\rangle  \\
\hline
\cdots  &\cdots & \cdots  \\
\hline
|k^2_{d_{2}-1,0}\rangle  &\cdots& |k^2_{d_{2}-1,d_{2}-1}\rangle  \\
\hline
\end{array}
\end{small}
\end{equation}

Assume $U_0,U_1,\dots,U_{d_1-1}$ are  unitary matrices  of order $d_2$ different from the identity matrix.  Define a unitary matrix $U$ of order $d_1d_2$:

\begin{equation}
U=\sum\limits_{i\in [d_1]}|i\rangle\langle i|\otimes U_i.\label{eqU}
\end{equation}
Let $\mathbb{I}$ be the identity matrix of order $d_1d_2$:

\begin{equation}
\mathbb{I}=\sum\limits_{i\in [d_1]}|i\rangle\langle i|\otimes \mathbb{I}_i,\label{eqI}
\end{equation}

\noindent where $\mathbb{I}_0,\dots, \mathbb{I}_{d_1-1}$ are the identity matrices of order $d_2$.

Define
\begin{equation}
\Phi=(|\Phi_{(i,m),(j,n)}\rangle)=(\tau |l^1_{i,j}\rangle\otimes |k^1_{m,n}\rangle)=(|l^1_{i,j}\rangle\otimes \tau_{l^1_{i,j}}|k^1_{m,n}\rangle),
\end{equation}
\begin{equation}
\Psi=(|\Psi_{(i,m),(j,n)}\rangle)=(\tau |l^2_{i,j}\rangle\otimes |k^2_{m,n}\rangle)=(|l^2_{i,j}\rangle\otimes \tau_{l^2_{i,j}}|k^2_{m,n}\rangle),
\end{equation}

\noindent where $\tau\in\{I,U\}$. Then $\Phi$, $\Psi$ can be written as the following squares:
\begin{equation}
\begin{small}
\setlength{\arraycolsep}{2.5pt}
\Phi=\begin{array}{|c|c|c|}
\hline
\tau |l^1_{0,0}\rangle \otimes  K^1  & \cdots & \tau |l^1_{0,d_1-1}\rangle \otimes  K^1   \\
\hline
\cdots  &\cdots & \cdots  \\
\hline
\tau |l^1_{d_1-1,0}\rangle \otimes  K^1   & \cdots & \tau |l^1_{d_1-1,d_1-1}\rangle \otimes  K^1   \\
\hline
\end{array}
\end{small}
\end{equation}
\begin{equation}
\begin{small}
\setlength{\arraycolsep}{2.5pt}
\Psi=\begin{array}{|c|c|c|}
\hline
\tau |l^2_{0,0}\rangle \otimes  K^2  & \cdots &\tau |l^2_{0,d_1-1}\rangle \otimes K^2  \\
\hline
\cdots  &\cdots & \cdots  \\
\hline
\tau |l^2_{d_1-1,0}\rangle \otimes  K^2   & \cdots &\tau |l^2_{d_1-1,d_1-1}\rangle \otimes  K^2   \\
\hline
\end{array}
\end{small}
\end{equation}

\noindent Any block $\tau |l^s_{i,j}\rangle \otimes K^s=\sum\limits_{i\in [d_1]}|i\rangle\langle i|\otimes \tau_i (|l^s_{i,j}\rangle \otimes K^s)=|l^s_{i,j}\rangle \otimes \tau_{l^s_{i,j}} K^s=\{|l^s_{i,j}\rangle \otimes \tau_{l^s_{i,j}} |k_{m,n}^s\rangle: m,n\in [d_2]\}$, where $\tau_{i}\in \{\mathbb{I}_{i},U_{i}\}$. Here, for each block in $\Phi$ and $\Psi$, the choice of $\tau$ from $\{\mathbb{I},U\}$ is independent.

It is easy to check that $\Phi$ and $\Psi$ are quantum Latin squares of dimension $d_1d_2$.  Furthermore, the set of vectors
$$\{|\Phi_{(i,m),(j,n)}\rangle\otimes |\Psi_{(i,m),(j,n)}\rangle: i,j\in [d_1],m,n\in [d_2] \} $$
 forms an orthonormal basis of the space $\mathbb{C}^{d_1d_2}\otimes \mathbb{C}^{d_1d_2}$, since

\vspace{0.1 cm}
\hspace{-0.4 cm}$ (|\Phi_{(i,m),(j,n)}\rangle\otimes |\Psi_{(i,m),(j,n)}\rangle, |\Phi_{(i',m'),(j',n')}\rangle\otimes |\Psi_{(i',m'),(j',n')}\rangle)$

\vspace{0.1 cm}\hspace{-0.3 cm}$=((|l^1_{i,j}\rangle \otimes \tau_{l^1_{i,j}} |k_{m,n}^1\rangle)\otimes (|l^2_{i,j}\rangle \otimes \tau_{l^2_{i,j}} |k_{m,n}^2\rangle),
(|l^1_{i',j'}\rangle \otimes \tau_{l^1_{i',j'}} |k_{m',n'}^1\rangle)
 \otimes (|l^2_{i',j'}\rangle \otimes $

 \hspace{0.2cm} $\tau_{l^2_{i',j'}} |k_{m',n'}^2\rangle))$

\vspace{0.1 cm}\hspace{-0.3 cm}$=(|l^1_{i,j}\rangle \otimes \tau_{l^1_{i,j}} |k_{m,n}^1\rangle, |l^1_{i',j'}\rangle \otimes \tau_{l^1_{i',j'}} |k_{m',n'}^1\rangle)(|l^2_{i,j}\rangle \otimes \tau_{l^2_{i,j}} |k_{m,n}^2\rangle,|l^2_{i',j'}\rangle \otimes \tau_{l^2_{i',j'}} |k_{m',n'}^2\rangle)$

\vspace{0.1 cm}\hspace{-0.3 cm}$=\langle l^1_{i,j}|l^1_{i',j'}\rangle\langle k^1_{m,n}|\tau^{\dagger}_{l^1_{i,j}}\tau_{l^1_{i',j'}} |k^1_{m',n'}\rangle
\langle l^2_{i,j}|l^2_{i',j'}\rangle\langle k^2_{m,n}|\tau^{\dagger}_{l^2_{i,j}}\tau_{l^2_{i',j'}} |k^2_{m',n'}\rangle$

\vspace{0.1 cm}\hspace{-0.3 cm}$=\langle k^1_{m,n}|\tau^{\dagger}_{l^1_{i,j}}\tau_{l^1_{i',j'}} |k^1_{m',n'}\rangle
\langle k^2_{m,n}|\tau^{\dagger}_{l^2_{i,j}}\tau_{l^2_{i',j'}} |k^2_{m',n'}\rangle\delta_{ii'}\delta_{jj'}$

\vspace{0.1 cm}\hspace{-0.3 cm}$=\langle k^1_{m,n}|k^1_{m',n'}\rangle\langle k^2_{m,n}|k^2_{m',n'}\rangle \delta_{ii'}\delta_{jj'}$

\vspace{0.1 cm}\hspace{-0.3 cm}$=\delta_{ii'}\delta_{jj'}\delta_{mm'}\delta_{nn'}.$

\vspace{0.1 cm}
\noindent So $\Phi$,$\Psi$ is a 2-MOQLS$(d_1d_2)$.\qed
\section{Proof of Construction~\ref{Sd1d2}}
\label{sec:AppB}

\noindent \p  Without loss of generality,  suppose $L$ is an HSOLS$(d_1^{n})$ on $\mathbb{Z}_{d_1n}$ with  holes  $S_0,S_1,\dots,S_{n-1}$, where $S_{i}=\{id_1,id_1+1,\dots,(i+1)d_1-1\}$ for any $i\in [n]$. Suppose $K$ is a  SOLS$(d_1)$ on $\mathbb{Z}_{d_1}$. Let $\Psi$, $\Phi$ be the corresponding classical HSOQLS$(d_1^{n})$ with  hole set $V=\{V_0,V_1,\dots,V_{n-1}\}$ and  classical SOQLS$(d_1)$ respectively, where the subspace $V_i=\Span\{|id_1\rangle,|id_1+1\rangle,\dots,|(i+1)d_1-1\rangle\}$ for any $i\in [n]$. Assume $U_0,U_1,\dots,U_{n-1}$ are  unitary matrices  of order $d_1$ different from the identity matrix.  Define a unitary matrix $U$ of order $d_1n$ as follows:
\vspace{-0.2cm}\begin{equation}
U=\sum\limits_{i\in [n]}|i\rangle\langle i|\otimes U_i.
\end{equation}

\vspace{-0.2cm} Filling each holes $V_i$ with $U(|i\rangle\otimes \Phi)=|i\rangle\otimes U_i\Phi$ for any $i\in [n]$, then the new square $\Psi'$ is a SOQLS$(d_1n)$.
\qed

\section{Proof of Construction~\ref{HMOQLS}}
\label{sec:AppC}

\noindent \p Suppose $\Psi^1=\{|\Psi^{1}_{i,j}\rangle\}$ and $\Psi^2=\{|\Psi^{2}_{i,j}\rangle\}$ is a pair of HMOQLS($h^n$) with hole set $V=\{V_1,V_2,\dots,V_n\}$ on $\mathbb{C}^{hn}$ and dim$V_s=h$ for $1\leq s\leq n$. Without loss of generality, assume the holes are in the diagonal line.  Put $\Phi^1=\{|\Phi^1_{l,k}\rangle\}$ and $\Phi^2=\{|\Phi^2_{l,k}\rangle\}$ to be a 2-MOQLS($m$) on $\mathbb{C}^m$.

Define two squares $\Psi$ and $\Phi$  on $\mathbb{C}^{hmn}$ with the hole set $V'=\{ V_1\otimes \mathbb{C}^m, V_2\otimes \mathbb{C}^m,\dots, V_n\otimes \mathbb{C}^m\}$. And let $\Psi=\{|\Psi_{(i,l),(j,k)}\rangle\}=\{|\Psi^1_{i,j}\rangle\otimes |\Phi^1_{l,k}\rangle\}$ and $\Phi=\{|\Phi_{(i,l),(j,k)}\rangle\}=\{|\Psi^2_{i,j}\rangle\otimes |\Phi^2_{l,k}\rangle\}$.  It is clear that $\Psi$ and $\Phi$ are both PIQLS$(hm)^n$s with the hole set $V'$. In addition, $\Psi$ and $\Phi$ are orthogonal. Since for any elements $|\Psi_{(i,l),(j,k)}\rangle$ and $|\Phi_{(i,l),(j,k)}\rangle$ in $\Psi$ and $\Phi$, $\{|\Psi_{(i,l),(j,k)}\rangle \otimes |\Phi_{(i,l),(j,k)}\rangle: i, j \in [hn],  l,k\in [m]\}$ is the orthonormal basis set of $(\mathbb{C}^{hmn} \otimes \mathbb{C}^{hmn} )\setminus \bigoplus_{i=1}^{n}((V_i\otimes \mathbb{C}^m)\otimes (V_i\otimes \mathbb{C}^m))$. In fact, for any $i,j,i',j' \in [hn]$, $l,k,l',k'\in [m]$,

\vspace{0.2 cm}\hspace{-0.2cm}
$
(|\Psi_{(i,l),(j,k)}\rangle \otimes |\Phi_{(i,l),(j,k)}\rangle, |\Psi_{(i',l'),(j',k')}\rangle \otimes |\Phi_{(i',l'),(j',k')}\rangle) $

\vspace{0.1 cm} $=((|\Psi^1_{i,j}\rangle\otimes |\Phi^1_{l,k}\rangle)\otimes (|\Psi^2_{i,j}\rangle\otimes |\Phi^2_{l,k}\rangle),
(|\Psi^1_{i',j'}\rangle\otimes |\Phi^1_{l',k'}\rangle)\otimes (|\Psi^2_{i',j'}\rangle\otimes |\Phi^2_{l',k'}\rangle))$

\vspace{0.1 cm}$ =(|\Psi^1_{i,j}\rangle\otimes|\Psi^2_{i,j}\rangle,|\Psi^1_{i',j'}\rangle\otimes|\Psi^2_{i',j'}\rangle)
(|\Phi^1_{l,k}\rangle\otimes|\Phi^2_{l,k}\rangle,|\Phi^1_{l',k'}\rangle\otimes|\Phi^2_{l',k'}\rangle)$

\vspace{0.1 cm}$=\delta_{ii'}\delta_{jj'}\delta_{ll'}\delta_{kk'}
$.

\vspace{0.2 cm}
Besides, if~$\Psi^1=\{|\Psi^{1}_{i,j}\rangle\}$, $\Psi^2=\{|\Psi^{2}_{i,j}\rangle\}$ is a pair of classical HMOQLS$(h^n)$, and $\Phi^1=\{|\Phi^1_{l,k}\rangle\}$, $\Phi^2=\{|\Phi^2_{l,k}\rangle\}$ is a pair of non-classical  2-MOQLS($m$). Then there exist some $(l_1,k_1)$,~$(l_2,k_2)$  satisfying~$|\langle\Phi^1_{l_1,k_1}|\Phi^1_{l_2,k_2}\rangle|\neq 0$ or~$\neq$ 1, where~$l_1,l_2,k_1,k_2\in [m]$. Thus for any~$i,j\in [hn]$, ~$|\langle\Psi_{(i,l_1),(j,k_1)}| \Psi_{(i,l_2),(j,k_2)}\rangle|=|\langle\Psi^1_{i,j}|\Psi^1_{i,j}\rangle \langle \Phi^1_{l_1,k_1}|\Phi^1_{l_2,k_2}\rangle|\\=|\langle \Phi^1_{l_1,k_1}|\Phi^1_{l_2,k_2}\rangle|\neq 0$ or~$\neq$ 1, so~$\Psi$ is a non-classical incomplete quantum Latin square, and the same with ~$\Phi$.
\qed
\section{Proof of Construction~\ref{HSOQLS}}
\label{sec:AppD}

\noindent \p Here we just prove that $\Phi$ defined by Eq.~(\ref{4.20}) is orthogonal with its conjugate transpose. In other words,  $\{|\Phi_{(i,l),(j,k)}\rangle \otimes |\Phi_{(j,k),(i,l)}\rangle^*: i, j \in [hn], l,k\in [m]\}$ is the orthonormal basis set of $(\mathbb{C}^{hmn} \otimes \mathbb{C}^{hmn} )\setminus \bigoplus_{i=1}^{n}((V_i\otimes \mathbb{C}^m)\otimes(V_i\otimes \mathbb{C}^m))$. In fact, for  any elements  $|\Phi_{(i,l),(j,k)}\rangle$ and $|\Phi_{(i',l'),(j',k')}\rangle$ in $\Phi$, assume $i\leq j$, and $i' \leq j'$, then

\vspace{0.2 cm}
$\hspace{0cm}
(|\Phi_{(i,l),(j,k)}\rangle \otimes |\Phi_{(j,k),(i,l)}\rangle^* , |\Phi_{(i',l'),(j',k')}\rangle\otimes |\Phi_{(j',k'),(i',l')}\rangle^*) $

\noindent \hspace{0.4cm}\vspace{0.1 cm} $=((|\Psi_{i,j}\rangle\otimes |\Phi^1_{l,k}\rangle)\otimes (|\Psi_{j,i}\rangle^*\otimes |\Phi^2_{l,k}\rangle),
(|\Psi_{i',j'}\rangle\otimes |\Phi^1_{l',k'}\rangle)\otimes (|\Psi_{j',i'}\rangle^*\otimes |\Phi^2_{l',k'}\rangle))$

\noindent \hspace{0.4cm}\vspace{0.1 cm} $=(|\Psi_{i,j}\rangle\otimes|\Psi_{j,i}\rangle^*,|\Psi_{i',j'}\rangle\otimes|\Psi_{j',i'}\rangle^*)
(|\Phi^1_{l,k}\rangle\otimes |\Phi^2_{l,k}\rangle,|\Phi^1_{l',k'}\rangle\otimes |\Phi^2_{l',k'}\rangle)$

\noindent \hspace{0.4cm}\vspace{0.1 cm} $=\delta_{ii'}\delta_{jj'}\delta_{ll'}\delta_{kk'}$.

\vspace{0.2 cm}In the same way, for any $i\leq j$ and $i' > j'$, $i>j$ and $i' \leq j'$, or $i> j$, and $i' > j'$, we always  get $
(|\Phi_{(i,l),(j,k)}\rangle \otimes |\Phi_{(j,k),(i,l)}\rangle^* , |\Phi_{(i',l'),(j',k')}\rangle\otimes |\Phi_{(j',k'),(i',l')}\rangle^*)
 =\delta_{ii'}\delta_{jj'}\delta_{ll'}\delta_{kk'}
$.
\qed

\section{Proof of Lemma~\ref{OA3}}
\label{sec:AppE}
\noindent \p Let $\{L^s:1\leq s \leq t\}$ be a set of $t$-MOLC($d$) with property (B) on $\mathbb{Z}_d$. Define a
$d^3\times (t+3)$ array $A =(a_{ijk})$ with rows  $(i,j,k, L^1_{i,j,k}, L^2_{i,j,k},\ldots,L^t_{i,j,k})$
for $i,j,k\in [d]$. Then $A$ is an orthogonal array OA$(d^3,t+3,d,3)$. This  process can be reversed to recover $t$ MOLS of order $d$ with property (B) from an OA$(d^3,t+3,d,3)$, by choosing the first three columns of the OA to index the rows, columns and files of the $t$ cubes. To show this more easily, we start with the OA.

Take any three columns $s_1,s_2,s_3$ of $A$ except for the first three columns and $s_1<s_2<s_3$.
We consider the following three cases: (1) $|\{s_1,s_2,s_3\}\cap \{1,2,3\}|=2$; (2) $|\{s_1,s_2,s_3\}\cap \{1,2,3\}|=1$; (3) $|\{s_1,s_2,s_3\}\cap \{1,2,3\}|=0$.

Case 1. When $|\{s_1,s_2,s_3\}\cap \{1,2,3\}|=2$. Then $(i,j,L^{s_3}_{i,j,k})$, $(j,k,L^{s_3}_{i,j,k})$ or $(i,k,L^{s_3}_{i,j,k})$ run through the full triples of $\mathbb{Z}_{d}^{\otimes3}$ if and only if  for any fixed $i$ and $j$, $j$ and $k$, or $i$ and $k$, the corresponding $L^{s_3}_{i,j,k}$ must run through the elements of $\mathbb{Z}_{d}$, i.e. $L^{s_3}$ is a Latin cube for any $1\leq s_3\leq t$.

Case 2. When $|\{s_1,s_2,s_3\}\cap \{1,2,3\}|=1$. Then $(i,L^{s_2}_{i,j,k},L^{s_3}_{i,j,k})$, $(j,L^{s_2}_{i,j,k},L^{s_3}_{i,j,k})$ or $(k,L^{s_2}_{i,j,k},L^{s_3}_{i,j,k})$  run through the full triples of $\mathbb{Z}_{d}^{\otimes3}$  if and only if for any fixed $i$, $j$, or $k$, the corresponding tuple $(L^{s_2}_{i,j,k},L^{s_3}_{i,j,k})$  run through the full tuples
 of $\mathbb{Z}_{d}^{\otimes2}$ , i.e. every corresponding planes of $L^{s_2},L^{s_3}$ are orthogonal for any $1\leq s_2< s_3\leq t$.

Case 3. When $|\{s_1,s_2,s_3\}\cap \{1,2,3\}|=0$. Then $(L^{s_1}_{i,j,k},L^{s_2}_{i,j,k},L^{s_3}_{i,j,k})$ run through the full triples of $\mathbb{Z}_{d}^{\otimes3}$  if and only if $L^{s_1}, L^{s_2},L^{s_3}$ are orthogonal for any $1\leq s_1<s_2< s_3\leq t$.
\qed

\section{Proof of Lemma~\ref{psi*}}
\label{sec:AppF}
\noindent \p Suppose $\Phi$, $\Psi$ and $\Upsilon$ are orthogonal quantum Latin cubes. Then
$\langle\Phi_{i,j,k}|\\ \Phi_{i',j',k'}\rangle\langle\Psi_{i,j,k}|\Psi_{i',j',k'}\rangle\langle\Upsilon_{i,j,k}|\Upsilon_{i',j',k'}\rangle=
\delta_{ii'}\delta_{jj'}\delta_{kk'}$  by Definition~\ref{MOQLC1} (1).
Thus $\langle\Phi_{i,j,k}|\Phi_{i',j',k'}\rangle=0$, $\langle\Psi_{i,j,k}|\Psi_{i',j',k'}\rangle=0,$ or $\langle\Upsilon_{i,j,k}|\Upsilon_{i',j',k'}\rangle=0$, for any $(i,j,k)\neq (i',j',k')$, else $\langle\Phi_{i,j,k}|\Phi_{i,j,k}\rangle=\langle\Psi_{i,j,k}|\Psi_{i,j,k}\rangle=\langle\Upsilon_{i,j,k}|\Upsilon_{i,j,k}\rangle=1.$  Since
 $\langle\Phi^*_{i,j,k}| \Phi^*_{i',j',k'}\rangle\\=\langle\Phi_{i,j,k}|\Phi_{i',j',k'}\rangle^*$, $\langle\Psi^*_{i,j,k}|\Psi^*_{i',j',k'}\rangle=$ $\langle\Psi_{i,j,k}|\Psi_{i',j',k'}\rangle^*$ and $0,1 \in \mathbb{R}$, so $\langle\Phi^*_{i,j,k}|\Phi^*_{i',j',k'}\rangle \\ \langle\Psi_{i,j,k} |\Psi_{i',j',k'}\rangle \langle\Upsilon_{i,j,k}|\Upsilon_{i',j',k'}\rangle=
\delta_{ii'}\delta_{jj'}\delta_{kk'}$ and $\langle\Phi^*_{i,j,k}|\Phi^*_{i',j',k'}\rangle\langle\Psi^*_{i,j,k}|\Psi^*_{i',j',k'}\rangle\\ \langle\Upsilon_{i,j,k}|\Upsilon_{i',j',k'}\rangle=
\delta_{ii'}\delta_{jj'}\delta_{kk'}$ hold.

On the other hand,  For each fixed $i$,$j$ or $k$, the corresponding planes of any two cubes coming from $\Phi^*$,$\Psi$ and $\Upsilon$ or $\Phi^*$,$\Psi^*$ and $\Upsilon$  can form a pair of orthogonal quantum Latin squares. Here we fix $i$  and consider $\Phi$, $\Psi$, then $\langle\Phi_{i,j,k}|\Phi_{i,j',k'}\rangle\langle\Psi_{i,j,k}|\Psi_{i,j',k'}\rangle\\=
\delta_{jj'}\delta_{kk'}$ by Definition~\ref{MOQLC1} (2).  ~So $\langle\Phi^*_{i,j,k}|\Phi^*_{i,j',k'}\rangle~\langle\Psi_{i,j,k}|\Psi_{i,j',k'}\rangle=
\delta_{jj'}\delta_{kk'}$ and $\langle\Phi^*_{i,j,k}|\Phi^*_{i,j',k'}\rangle~\langle\Psi^*_{i,j,k}|\\ \Psi^*_{i,j',k'}\rangle=
\delta_{jj'}\delta_{kk'}$. Moreover it's true for other cases.

Thus, $\Phi^*$, $\Psi$ and $\Upsilon$ as well as $\Phi^*$, $\Psi^*$ and $\Upsilon$ are two triples of orthogonal quantum Latin cubes. The converse then follows since $(\Phi^*)^*=\Phi$, $(\Psi^*)^*=\Psi$.
\qed
\section{Proof of Construction~\ref{cd1d2}}
\label{sec:AppG}
\noindent \p  Suppose $\mathbb{C}^{d_1}=\Span \{|0\rangle,|1\rangle,\dots,|d_1-1\rangle\}$ and $\mathbb{C}^{d_2}=\Span\{|0\rangle,|1\rangle,\dots,|d_2-1\rangle\}$. Then $\mathbb{C}^{d_1d_2}\simeq \mathbb{C}^{d_1}\otimes\mathbb{C}^{d_2}=\Span\{|i\rangle \otimes |j\rangle: i\in [d_1], j\in [d_2]
\}=\Span\{|0\rangle,|1\rangle,\dots,\\|d_1d_2-1\rangle\}$.

Let $l^s=(l^s_{i,j,k})_{d_1\times d_1\times d_1}$ and $k^s=(k^s_{f,g,h})_{d_2\times d_2\times d_2}$,  $1 \leq s\leq 3$, are 3-MOLC$(d_1)$ and 3-MOLC$(d_2)$ with the Property (B) respectively. Then put $L^s=\{|l^s_{i,j,k}\rangle: i,j,k \in [d_1]\}$, $K^s=\{|k^s_{f,g,h}\rangle: f,g,h \in [d_2]\}$ to be the corresponding classical quantum Latin cubes of $l^s$ and $k^s$, $1 \leq s\leq 3$. Define a unitary matrix $U$ and identity matrix $\mathbb{I}$ of order $d_1d_2$ as Eqs.~(\ref{eqU})--(\ref{eqI}).

Let
$$\Phi=(|\Phi_{(i,f),(j,g),(k,h)}\rangle)=(\tau |l^1_{i,j,k}\rangle\otimes |k^1_{f,g,h}\rangle)=(|l^1_{i,j,k}\rangle\otimes \tau_{l^1_{i,j,k}}|k^1_{f,g,h}\rangle),$$
$$\Psi=(|\Psi_{(i,f),(j,g),(k,h)}\rangle)=(\tau |l^2_{i,j,k}\rangle\otimes |k^2_{f,g,h}\rangle)=(|l^2_{i,j,k}\rangle\otimes \tau_{l^2_{i,j,k}}|k^2_{f,g,h}\rangle),$$
$$\Upsilon=(|\Upsilon_{(i,f),(j,g),(k,h)}\rangle)=(\tau |l^3_{i,j,k}\rangle\otimes |k^3_{f,g,h}\rangle)=(|l^3_{i,j,k}\rangle\otimes \tau_{l^3_{i,j,k}}|k^3_{f,g,h}\rangle),$$
 where
$\tau \in \{\mathbb{I},U\}$,  $\tau_{i}\in\{\mathbb{I}_{i},U_{i}\}$, $i,j,k \in [d_1]$ and $f,g,h \in [d_2]$. Here we choose $\tau$ from $\{\mathbb{I},U\}$ for each block independently  in $\Phi$, $\Psi$ or $\Upsilon$.

It is easy to see $\Phi$, $\Psi$ and $\Upsilon$ are quantum Latin cubes. Moreover,  they are orthogonal.

(1) The set of vectors
$$\{|\Phi_{(i,f),(j,g),(k,h)}\rangle\otimes |\Psi_{(i,f),(j,g),(k,h)}\rangle \otimes |\Upsilon_{(i,f),(j,g),(k,h)}\rangle : i,j,k\in [d_1],f,g,h\in [d_2] \} $$
forms an orthonormal basis of the space $\mathbb{C}^{d_1d_2}\otimes \mathbb{C}^{d_1d_2}\otimes \mathbb{C}^{d_1d_2}$.
By the definition of  orthogonal classical Latin cube, we  get

\vspace{0.2 cm}\hspace{-0.6cm}
$(|\Phi_{(i,f),(j,g),(k,h)}\rangle\otimes |\Psi_{(i,f),(j,g),(k,h)}\rangle \otimes |\Upsilon_{(i,f),(j,g),(k,h)}\rangle, |\Phi_{(i',f'),(j',g'),(k',h')}\rangle\otimes$

\vspace{0.1 cm}\hspace{-0.2cm}$ |\Psi_{(i',f'),(j',g'),(k',h')}\rangle \otimes |\Upsilon_{(i',f'),(j',g'),(k',h')}\rangle)$

\vspace{0.1 cm}\hspace{-0.4cm}$=\langle\Phi_{(i,f),(j,g),(k,h)}|\Phi_{(i',f'),(j',g'),(k',h')}\rangle\langle\Psi_{(i,f),(j,g),(k,h)}
|\Psi_{(i',f'),(j',g'),(k',h')}\rangle$

\vspace{0.1 cm}\hspace{0cm}$\langle\Upsilon_{(i,f),(j,g),(k,h)}|\Upsilon_{(i',f'),(j',g'),(k',h')}\rangle$

\vspace{0.1 cm}\hspace{-0.4cm}$=\langle l^1_{i,j,k}|l^1_{i',j',k'}\rangle\langle k^1_{f,g,h}|\tau^{\dagger}_{l^1_{i,j,k}}\tau_{l^1_{i',j',k'}}|k^1_{f',g',h'}\rangle
\langle l^2_{i,j,k}|l^2_{i',j',k'}\rangle\langle k^2_{f,g,h}|\tau^{\dagger}_{l^2_{i,j,k}}\tau_{l^2_{i',j',k'}}|k^2_{f',g',h'}\rangle$

\vspace{0.1 cm}\hspace{0cm}$\langle l^3_{i,j,k}|
l^3_{i',j',k'}\rangle\langle k^3_{f,g,h}|\tau^{\dagger}_{l^3_{i,j,k}}\tau_{l^3_{i',j',k'}}|k^3_{f',g',h'}\rangle$

\vspace{0.1 cm}\hspace{-0.4cm}$=\langle k^1_{f,g,h}|\tau^{\dagger}_{l^1_{i,j,k}}\tau_{l^1_{i',j',k'}}|k^1_{f',g',h'}\rangle
\langle k^2_{f,g,h}|\tau^{\dagger}_{l^2_{i,j,k}}\tau_{l^2_{i',j',k'}}|k^2_{f',g',h'}\rangle
\langle k^3_{f,g,h}|\tau^{\dagger}_{l^3_{i,j,k}}\tau_{l^3_{i',j',k'}}|k^3_{f',g',h'}\rangle $

\hspace{0cm}$ \delta_{ii'}\delta_{jj'}\delta_{kk'}$

\vspace{0.1 cm}\hspace{-0.4cm}$=\langle k^1_{f,g,h}|k^1_{f',g',h'}\rangle\langle k^2_{f,g,h}|k^2_{f',g',h'}\rangle\langle k^3_{f,g,h}|k^3_{f',g',h'}\rangle\delta_{ii'}\delta_{jj'}\delta_{kk'}$

\vspace{0.1 cm}\hspace{-0.4cm}$= \delta_{ii'}\delta_{jj'}\delta_{kk'} \delta_{ff'}\delta_{gg'}\delta_{hh'}$.

\vspace{0.2 cm}
(2) For each fixed $(i,f)$,$(j,g)$ or $(k,h)$, the corresponding planes of any two cubes coming from $\Phi$,$\Psi$ and $\Upsilon$  form a pair of orthogonal quantum Latin squares. Here we fix $(k,h)$, and show that the two  corresponding planes of $\Phi$ and $\Psi$ is a pair of orthogonal quantum Latin squares. That is to say for fixed $(k,h)$,  the set of vectors
$\{|\Phi_{(i,f),(j,g),(k,h)}\rangle\otimes |\Psi_{(i,f),(j,g),(k,h)}\rangle: i,j\in [d_1],f,g\in [d_2] \} $ forms an orthonormal basis of the space $\mathbb{C}^{d_1d_2}\otimes \mathbb{C}^{d_1d_2}$. It is true, because by the property (B) in Lemma~\ref{OA3}, we  get

\vspace{0.2 cm}\hspace{-0.2cm}
$(|\Phi_{(i,f),(j,g),(k,h)}\rangle\otimes |\Psi_{(i,f),(j,g),(k,h)}\rangle, |\Phi_{(i',f'),(j',g'),(k,h)}\rangle\otimes|\Phi_{(i',f'),(j',g'),(k,h)}\rangle)$

\vspace{0.1 cm}$=\langle\Phi_{(i,f),(j,g),(k,h)}|\Phi_{(i',f'),(j',g'),(k,h)}\rangle\langle\Psi_{(i,f),(j,g),(k,h)}
|\Psi_{(i',f'),(j',g'),(k,h)}\rangle$

\vspace{0.1 cm}$=\langle l^1_{i,j,k}|l^1_{i',j',k}\rangle\langle k^1_{f,g,h}|\tau^{\dagger}_{l^1_{i,j,k}}\tau_{l^1_{i',j',k}}|k^1_{f',g',h}\rangle
\langle l^2_{i,j,k}|l^2_{i',j',k}\rangle\langle k^2_{f,g,h}|\tau^{\dagger}_{l^2_{i,j,k}}\tau_{l^2_{i',j',k}}|k^2_{f',g',h}\rangle$

\vspace{0.1 cm}$=\langle k^1_{f,g,h}|\tau^{\dagger}_{l^1_{i,j,k}}\tau_{l^1_{i',j',k}}|k^1_{f',g',h}\rangle
\langle k^2_{f,g,h}|\tau^{\dagger}_{l^2_{i,j,k}}\tau_{l^2_{i',j',k}}|k^2_{f',g',h}\rangle \delta_{ii'}\delta_{jj'}$

\vspace{0.1 cm}$=\langle k^1_{f,g,h}|k^1_{f',g',h}\rangle\langle k^2_{f,g,h}|k^2_{f',g',h}\rangle \delta_{ii'}\delta_{jj'}$

\vspace{0.1 cm}$= \delta_{ii'}\delta_{jj'}\delta_{ff'}\delta_{gg'}
$.

\vspace{0.2cm}
Moreover it is true for the other cases. So $\Phi$, $\Psi$ and $\Upsilon$ is a triple of orthogonal quantum Latin cubes.
\qed
\section{Proof of Proposition~\ref{GMOQLC}}
\label{sec:AppH}

\noindent \p Suppose that $|\Phi\rangle$ is the sum of the $d^3$ states in the QOA$(d^3,t+3,d,3)$.   Since $|\Phi\rangle$ can  produce a 3-uniform state, we choose the first three subsystems, namely, $i$, $j$, $k$, then
\begin{equation}
|\Phi\rangle=\sum^{d-1}_{i,j,k=0}|ijk\rangle\otimes|\Phi_{i,j,k}\rangle.
\end{equation}

Actually, $\{|\Phi_{i,j,k}\rangle\}$ is the set of arrangements of $t$-GMOQLC$(d)$, where $i,j,k\in [d]$ are the indexes of the rows, columns and files of the $t$-GMOQLC$(d)$.
To show this, we take an arbitrary subset $S \subseteq \{1,2,\dots,t+3\}$ with $|S|=3$. Consider the following four cases: (1) $|S\cap\{1,2,3\}|=3$; (2) $|S\cap \{1,2,3\}|=2$; (3) $|S\cap \{1,2,3\}|=1$; (4) $|S\cap \{1,2,3\}|=0$.

Case 1. When $|S\cap\{1,2,3\}|=3$, we have
\begin{eqnarray*}
\rho_{S}&=Tr_{4,5,\dots,t+3}\sum_{i,j,k,i',j',k'\in [d]}|ijk\rangle\otimes|\Phi_{i,j,k}\rangle \langle i'j'k'|\otimes\langle\Phi_{i',j',k'}|\\
&=\sum_{i,j,k,i',j',k'\in [d]}|ijk\rangle\langle i'j'k'| \langle\Phi_{i',j',k'}|\Phi_{i,j,k}\rangle.
\end{eqnarray*}
\noindent So, $\rho_{S}=\sum\limits_{i,j,k,i',j',k'\in [d]}|ijk\rangle\langle i'j'k'| \langle\Phi_{i',j',k'}|\Phi_{i,j,k}\rangle= \mathbb{I}_{d^{3}}$ if and only if  Eq.~(\ref{eq1'}) holds.

Case 2. When $|S\cap \{1,2,3\}|=2$. Suppose $S\cap \{1,2,3\}=\{1,2\}$, and $\{l_1,l_2,\dots, l_{t-1}\}\\ \cap \{1,2,3\}=\emptyset$, then we have
\begin{eqnarray*}
\rho_{S}&=Tr_{3,l_1,l_2,\dots, l_{t-1}}\sum_{i,j,k,i',j',k'\in [d]}|ijk\rangle\otimes|\Phi_{i,j,k}\rangle \langle i'j'k'|\otimes\langle\Phi_{i',j',k'}|\\
&=\sum_{i,j,i',j'\in [d]}|ij\rangle\langle i'j'| \otimes \sum_{k\in [d]} Tr_{l_1,l_2,\dots, l_{t-1}} |\Phi_{i,j,k}\rangle \langle\Phi_{i',j',k}|.
\end{eqnarray*}
\noindent So, $\sum\limits_{i,j,i',j'\in [d]}|ij\rangle\langle i'j'| \otimes \sum\limits_{k\in [d]} Tr_{l_1,l_2,\dots, l_{t-1}} |\Phi_{i,j,k}\rangle \langle\Phi_{i',j',k}|= \mathbb{I}_{d^{3}}$ if and only if  Eq.~(\ref{eq4'}) holds.
If $S\cap \{1,2,3\}=\{1,3\}$, or $S\cap \{1,2,3\}=\{2,3\}$, in that case, they hold true if and only if  Eqs.~(\ref{eq3'}) or~(\ref{eq2'}) holds true, respectively.

Case 3. When $|S\cap \{1,2,3\}|=1$. Suppose $S\cap \{1,2,3\}=\{1\}$, and $\{l_1,l_2,\dots, l_{t-2}\}\cap \{1,2,3\}=\emptyset$, then we have
\begin{eqnarray*}
\rho_{S}&=Tr_{2,3,l_1,l_2,\dots, l_{t-2}}\sum_{i,j,k,i',j',k'\in [d]}|ijk\rangle\otimes|\Phi_{i,j,k}\rangle \langle i'j'k'|\otimes\langle\Phi_{i',j',k'}|\\
&=\sum_{i,i'\in [d]}|i\rangle\langle i'| \otimes \sum_{j,k\in [d]} Tr_{l_1,l_2,\dots, l_{t-1}} |\Phi_{i,j,k}\rangle \langle\Phi_{i',j,k}|.
\end{eqnarray*}
\noindent So, $\rho_{S}= \sum\limits_{i,i'\in [d]}|i\rangle\langle i'| \otimes \sum\limits_{j,k\in [d]} Tr_{l_1,l_2,\dots, l_{t-1}} |\Phi_{i,j,k}\rangle \langle\Phi_{i',j,k}|= \mathbb{I}_{d^{3}}$ if and only if  Eq.~(\ref{eq6'}) holds.
If $S\cap \{1,2,3\}=\{2\}$, or $S\cap \{1,2,3\}=\{3\}$, they hold true if and only if Eqs.~(\ref{eq7'}) or~(\ref{eq5'}) holds, respectively.

Case 4. When $|S\cap \{1,2,3\}|=0$, we have
\begin{eqnarray*}
\rho_{S}&=Tr_{1,2,3,l_1,l_2,\dots, l_{t-3}}\sum_{i,j,k,i',j',k'\in [d]}|ijk\rangle\otimes|\Phi_{i,j,k}\rangle \langle i'j'k'|\otimes\langle\Phi_{i',j',k'}|\\
&=\sum_{i,j,k\in [d]} Tr_{l_1,l_2,\dots, l_{t-2}} |\Phi_{i,j,k}\rangle \langle\Phi_{i,j,k}|.
\end{eqnarray*}
\noindent So, $\rho_{S}= \mathbb{I}_{d^{3}}$ if and only if  Eq.~(\ref{eq8'}) holds.
\qed

\section{Proof of  Example~\ref{Q16}}
\label{sec:AppI}

\noindent \p Define

$$U=\sum\limits_{i\in [4]}|i\rangle\langle i|\otimes U_i,$$

\noindent where
\begin{eqnarray}
  \begin{small}
\setlength{\arraycolsep}{2.5pt}
~~~~~~~ U_0=\frac{1}{2}\left(
       \begin{array}{cccc}
1 & 1& 1&1\\
1 &-1&1&-1\\
1 &1& -1&-1\\
1 &-1& -1&1\\
\end{array}
        \right),\hspace{0.9 cm}
        U_1=\left(
        \begin{array}{cccc}
\frac{1 }{4}& -\frac{\sqrt{-3}}{4}& -\frac{\sqrt{-3}}{4}&-\frac{3}{4}\\
\frac{\sqrt{3}}{4} &\frac{\sqrt{-1}}{4}&-\frac{3\sqrt{-1}}{4}&\frac{\sqrt{3}}{4}\\
\frac{\sqrt{3}}{4} &-\frac{3\sqrt{-1}}{4}& \frac{\sqrt{-1}}{4}&\frac{\sqrt{3}}{4}\\
\frac{3}{4} &\frac{\sqrt{-3}}{4}& \frac{\sqrt{-3}}{4}&-\frac{1}{4}\\
\end{array}
        \right),
        \end{small}\label{00}
   \\ \nonumber
      \begin{small}
\setlength{\arraycolsep}{2.5pt}
   U_2=\left(
        \begin{array}{cccc}
\frac{1}{2} &\frac{\sqrt{-1}}{2}& \frac{\sqrt{-1}}{2}&-\frac{1}{2}\\
\frac{\sqrt{-1}}{2} &\frac{1}{2}&-\frac{1}{2}&\frac{\sqrt{-1}}{2}\\
\frac{\sqrt{-1}}{2} &-\frac{1}{2}& \frac{1}{2}&\frac{\sqrt{-1}}{2}\\
-\frac{1}{2} &\frac{\sqrt{-1}}{2}& \frac{\sqrt{-1}}{2}&\frac{1}{2}\\
\end{array}
        \right), \hspace{0.4 cm}
        U_3=\left(
       \begin{array}{cccc}
\frac{\sqrt{2}}{4} &\frac{\sqrt{-2}}{4}&-\frac{\sqrt{-6}}{4}&\frac{\sqrt{6}}{4}\\
\frac{\sqrt{-2}}{4} &\frac{\sqrt{2}}{4}&\frac{\sqrt{6}}{4}&-\frac{\sqrt{-6}}{4}\\
\frac{\sqrt{6}}{4}&\frac{\sqrt{-6}}{4}& \frac{\sqrt{-2}}{4}&-\frac{\sqrt{2}}{4}\\
\frac{\sqrt{-6}}{4} &\frac{\sqrt{6}}{4}& -\frac{\sqrt{2}}{4}&\frac{\sqrt{-2}}{4}\\
\end{array}
\right).
\end{small}
\end{eqnarray}

\vspace{0.2 cm}
Let $\Psi$ be a classical HSOQLS$(4^4)$ with holes~$V_i=span\{|4i\rangle,|4i+1\rangle,|4i+2\rangle,|4i+3\rangle\}$ on $\mathbb{C}^{16}$, $i \in [4]$, as follows.

\vspace{0.2 cm}
\begin{equation*}
\begin{small}
\setlength{\arraycolsep}{0.2 pt}
\begin{array}{lc}
\mbox{}&
\begin{array}{cccccccccccccccc}~~|0\rangle &~~|1\rangle&~ ~|2\rangle~&  ~|3\rangle& ~|4\rangle & ~~|5\rangle &~~|6\rangle & ~~|7\rangle   &~|8\rangle & ~|9\rangle & |10\rangle& |11\rangle& |12\rangle & |13\rangle & |14\rangle & |15\rangle \end{array}\\
\begin{array}{c}|0\rangle \\|1\rangle\\ |2\rangle\\  |3\rangle\\|4\rangle\\|5\rangle \\ |6\rangle \\ |7\rangle \\|8\rangle\\ |9\rangle \\ |10\rangle\\ |11\rangle\\|12\rangle \\ |13\rangle \\|14\rangle \\ |15\rangle \end{array}&
\begin{array}{|c|c|c|c|c|c|c|c|c|c|c|c|c|c|c|c|}
\hline
 &&&&	|12\rangle &	|13\rangle &	|14\rangle &	 |15\rangle &	|4\rangle &	|5\rangle &	|6\rangle &	|7\rangle &	 |8\rangle &	 |9\rangle &	 |10\rangle &	|11\rangle \\
\hline
&&& &	|14\rangle &	|15\rangle &	|12\rangle &	 |13\rangle &	|6\rangle &	|7\rangle &	 |4\rangle &	 |5\rangle &	|10\rangle &	 |11\rangle &	 |8\rangle &	|9\rangle \\
\hline
&&&&	|15\rangle &	|14\rangle &	|13\rangle &	 |12\rangle &	|7\rangle &	|6\rangle &	 |5\rangle &	 |4\rangle &	|11\rangle &	 |10\rangle &	 |9\rangle &	|8\rangle \\
\hline
&&& &	|13\rangle &	|12\rangle &	|15\rangle &	 |14\rangle &	|5\rangle &	|4\rangle &	|7\rangle &	 |6\rangle &	 |9\rangle &	 |8\rangle &	 |11\rangle &	|10\rangle \\
\hline
|8\rangle &	|9\rangle &	|10\rangle &	|11\rangle &	&&& &	|12\rangle &	|13\rangle &	 |14\rangle &	 |15\rangle &	|0\rangle &	 |1\rangle &	 |2\rangle &	|3\rangle \\
\hline
|11\rangle &	|10\rangle &	|9\rangle &	|8\rangle &	&&& &	|14\rangle &	|15\rangle &	 |12\rangle &	 |13\rangle &	|2\rangle &	 |3\rangle &	 |0\rangle &	|1\rangle \\
\hline
|9\rangle &	|8\rangle &	|11\rangle &	|10\rangle &	&&& &	|15\rangle &	|14\rangle &	 |13\rangle &	 |12\rangle &	|3\rangle &	 |2\rangle &	 |1\rangle &	|0\rangle \\
\hline
|10\rangle &	|11\rangle &	|8\rangle &	|9\rangle &	&&& &	|13\rangle &	|12\rangle &	 |15\rangle &	 |14\rangle &	|1\rangle &	 |0\rangle &	 |3\rangle &	|2\rangle \\
\hline
|12\rangle &	|13\rangle &	|14\rangle &	|15\rangle &	|0\rangle &	|1\rangle &	|2\rangle &	|3\rangle &	&&& &	 |4\rangle &	 |5\rangle &	 |6\rangle &	 |7\rangle \\
\hline
|15\rangle &	|14\rangle &	|13\rangle &	|12\rangle &	|3\rangle &	|2\rangle &	|1\rangle &	|0\rangle &&&& &	 |6\rangle &	 |7\rangle &	 |4\rangle &	 |5\rangle \\
\hline
|13\rangle &	|12\rangle &	|15\rangle &	|14\rangle &	|1\rangle &	|0\rangle &	|3\rangle &	|2\rangle &	&&& &	 |7\rangle &	 |6\rangle &	 |5\rangle &	 |4\rangle \\
\hline
|14\rangle &	|15\rangle &	|12\rangle &	|13\rangle &	|2\rangle &	|3\rangle &	|0\rangle &	|1\rangle &&&& &	 |5\rangle &	 |4\rangle &	 |7\rangle &	 |6\rangle \\
\hline
|4\rangle &	|5\rangle &	|6\rangle &	|7\rangle &	|8\rangle &	|9\rangle &	|10\rangle &	|11\rangle &	|0\rangle &	 |1\rangle &	 |2\rangle &	 |3\rangle &	&&& \\
\hline
|7\rangle &	|6\rangle &	|5\rangle &	|4\rangle &	|11\rangle &	|10\rangle &	|9\rangle &	|8\rangle &	|3\rangle &	 |2\rangle &	 |1\rangle &	 |0\rangle &	&&& \\
\hline
|5\rangle &	|4\rangle &	|7\rangle &	|6\rangle &	|9\rangle &	|8\rangle &	|11\rangle &	|10\rangle &	|1\rangle &	 |0\rangle &	 |3\rangle &	 |2\rangle &&&& \\
\hline
|6\rangle &	|7\rangle &	|4\rangle &	|5\rangle &	|10\rangle &	|11\rangle &	|8\rangle &	|9\rangle &	|2\rangle &	 |3\rangle &	 |0\rangle &	 |1\rangle &	&&& \\
\hline
\end{array}
\end{array}
\end{small}
\end{equation*}

\vspace{0.2 cm}
Here is a classical SOQLS(4) on $\mathbb{C}^4$:

$$\begin{array}{lc}
\mbox{}&
\Phi=\begin{array}{|c|c|c|c|}
\hline
|0\rangle &  |1\rangle  &|2\rangle& |3\rangle \\
\hline
|3\rangle & |2\rangle & |1\rangle &|0\rangle \\
\hline
|1\rangle &|0\rangle   & |3\rangle & |2\rangle  \\
\hline
|2\rangle & |3\rangle  &|0\rangle &  |1\rangle \\
\hline
\end{array}
\end{array}$$

By the process in the proof of Construction~\ref{Sd1d2}, let~$\Phi_i=U(|i\rangle\otimes \Phi)=|i\rangle\otimes U_i\Phi \in \mathbb{C}^{16}$, ~$ i\in [n]$.  We  get a SOQLS(16) $\Psi'$ by filling the holes $V_i$ with $\Phi_i$ in $\Psi$.

\vspace{0.2 cm}

\noindent\begin{small}
\setlength{\arraycolsep}{0.5 pt}
$\begin{array}{lc}
\mbox{}&
\begin{array}{cccccccccccccccc}~|0\rangle &~~~|1\rangle&~~|2\rangle& ~~|3\rangle&~~~|4\rangle &~~|5\rangle &~~|6\rangle &~~|7\rangle   &~~~|8\rangle &~~|9\rangle &~~~|10\rangle& ~~|11\rangle& ~~~|12\rangle & ~~|13\rangle &~ ~~|14\rangle &~ ~~|15\rangle \end{array}\\
\begin{array}{c}|0\rangle \\|1\rangle\\ |2\rangle\\  |3\rangle\\|4\rangle\\|5\rangle \\ |6\rangle \\ |7\rangle \\|8\rangle\\ |9\rangle \\ |10\rangle\\ |11\rangle\\|12\rangle \\ |13\rangle \\|14\rangle \\ |15\rangle \end{array}&
\begin{array}{|c|c|c|c|c|c|c|c|c|c|c|c|c|c|c|c|}
\hline
{\color{red}U|0\rangle}&	{\color{red}U|1\rangle}&	{\color{red}U|2\rangle} &	{\color{red}U|3\rangle}&	 |12\rangle &	|13\rangle &	 |14\rangle &	 |15\rangle &	|4\rangle &	|5\rangle &	|6\rangle &	|7\rangle &	 |8\rangle &	|9\rangle &	 |10\rangle &	|11\rangle \\
\hline
{\color{red}U|3\rangle} &	{\color{red}U|2\rangle} &	{\color{red}U|1\rangle} &	{\color{red}U|0\rangle} &	 |14\rangle &	|15\rangle &	 |12\rangle &	 |13\rangle &	|6\rangle &	|7\rangle &	 |4\rangle &	|5\rangle &	 |10\rangle &	|11\rangle &	 |8\rangle &	 |9\rangle \\
\hline
{\color{red}U|1\rangle} &	{\color{red}U|0\rangle} &	{\color{red}U|3\rangle} &	{\color{red}U|2\rangle} &	 |15\rangle &	|14\rangle &	 |13\rangle &	 |12\rangle &	|7\rangle &	|6\rangle &	 |5\rangle &	|4\rangle &	 |11\rangle &	|10\rangle &	 |9\rangle &	 |8\rangle \\
\hline
{\color{red}U|2\rangle} &	{\color{red}U|3\rangle} &	{\color{red}U|0\rangle} &	{\color{red}U|1\rangle} &	 |13\rangle &	|12\rangle &	 |15\rangle &	 |14\rangle &	|5\rangle &	|4\rangle &	|7\rangle &	 |6\rangle &	 |9\rangle &	|8\rangle &	 |11\rangle &	 |10\rangle \\
\hline
|8\rangle &	|9\rangle &	|10\rangle &	|11\rangle &	{\color{red}U|4\rangle}&	{\color{red}U|5\rangle}&	 {\color{red}U|6\rangle} &	 {\color{red}U|7\rangle}&	|12\rangle &	|13\rangle &	 |14\rangle &	|15\rangle &	 |0\rangle &	|1\rangle &	 |2\rangle &	 |3\rangle \\
\hline
|11\rangle &	|10\rangle &	|9\rangle &	|8\rangle &{\color{red}U|7\rangle}&	{\color{red}U|6\rangle}&	 {\color{red}U|5\rangle} &	 {\color{red}U|4\rangle}&	|14\rangle &	|15\rangle &	 |12\rangle &	|13\rangle &	 |2\rangle &	|3\rangle &	 |0\rangle &	 |1\rangle \\
\hline
|9\rangle &	|8\rangle &	|11\rangle &	|10\rangle &	{\color{red}U|5\rangle}&	{\color{red}U|4\rangle}&	 {\color{red}U|7\rangle} &	 {\color{red}U|6\rangle}&	|15\rangle &	|14\rangle &	 |13\rangle &	|12\rangle &	 |3\rangle &	|2\rangle &	 |1\rangle &	 |0\rangle \\
\hline
|10\rangle &	|11\rangle &	|8\rangle &	|9\rangle &	{\color{red}U|6\rangle}&	{\color{red}U|7\rangle}&	 {\color{red}U|4\rangle} &	 {\color{red}U|5\rangle}&	|13\rangle &	|12\rangle &	 |15\rangle &	|14\rangle &	 |1\rangle &	|0\rangle &	 |3\rangle &	 |2\rangle \\
\hline
|12\rangle &	|13\rangle &	|14\rangle &	|15\rangle &	|0\rangle &	|1\rangle &	|2\rangle &	|3\rangle &	 {\color{red}U|8\rangle}&	 {\color{red}U|9\rangle}&	{\color{red}U|10\rangle} &	{\color{red}U|11\rangle}&	 |4\rangle &	|5\rangle &	 |6\rangle &	 |7\rangle \\
\hline
|15\rangle &	|14\rangle &	|13\rangle &	|12\rangle &	|3\rangle &	|2\rangle &	|1\rangle &	|0\rangle &{\color{red}U|11\rangle}&	 {\color{red}U|10\rangle}&	{\color{red}U|9\rangle} &	{\color{red}U|8\rangle}&	 |6\rangle &	|7\rangle &	 |4\rangle &	 |5\rangle \\
\hline
|13\rangle &	|12\rangle &	|15\rangle &	|14\rangle &	|1\rangle &	|0\rangle &	|3\rangle &	|2\rangle &	 {\color{red}U|9\rangle}&	 {\color{red}U|8\rangle}&	{\color{red}U|11\rangle} &	{\color{red}U|10\rangle}&	 |7\rangle &	|6\rangle &	 |5\rangle &	 |4\rangle \\
\hline
|14\rangle &	|15\rangle &	|12\rangle &	|13\rangle &	|2\rangle &	|3\rangle &	|0\rangle &	|1\rangle &{\color{red}U|10\rangle}&	 {\color{red}U|11\rangle}&	{\color{red}U|8\rangle} &	{\color{red}U|9\rangle}&	 |5\rangle &	|4\rangle &	 |7\rangle &	 |6\rangle \\
\hline
|4\rangle &	|5\rangle &	|6\rangle &	|7\rangle &	|8\rangle &	|9\rangle &	|10\rangle &	|11\rangle &	|0\rangle &	 |1\rangle &	 |2\rangle &	 |3\rangle &	{\color{red}U|12\rangle}&	{\color{red}U|13\rangle}&	 {\color{red}U|14\rangle} &	{\color{red}U|15\rangle} \\
\hline
|7\rangle &	|6\rangle &	|5\rangle &	|4\rangle &	|11\rangle &	|10\rangle &	|9\rangle &	|8\rangle &	|3\rangle &	 |2\rangle &	 |1\rangle &	 |0\rangle &	{\color{red}U|15\rangle}&	{\color{red}U|14\rangle}&	 {\color{red}U|13\rangle} &	{\color{red}U|12\rangle}\\
\hline
|5\rangle &	|4\rangle &	|7\rangle &	|6\rangle &	|9\rangle &	|8\rangle &	|11\rangle &	|10\rangle &	|1\rangle &	 |0\rangle &	 |3\rangle &	 |2\rangle &{\color{red}U|13\rangle}&	{\color{red}U|12\rangle}&	 {\color{red}U|15\rangle} &	{\color{red}U|14\rangle} \\
\hline
|6\rangle &	|7\rangle &	|4\rangle &	|5\rangle &	|10\rangle &	|11\rangle &	|8\rangle &	|9\rangle &	|2\rangle &	 |3\rangle &	 |0\rangle &	 |1\rangle &	{\color{red}U|14\rangle}&	{\color{red}U|15\rangle}&	 {\color{red}U|12\rangle} &	{\color{red}U|13\rangle} \\
\hline
\end{array}
\end{array}.$
\end{small}

\vspace{0.2 cm}
Furthermore, $\Psi'$ is a non-classical quantum Latin square, since  $|\langle\Psi'_{0,8}|\Psi'_{4,5}\rangle|=|\langle 4|U|5\rangle|=|\frac{-\sqrt{-3}}{4}|\neq 0$ or $\neq$ 1, where we set $|4\rangle=|1\rangle\otimes|0\rangle$ and $|5\rangle=|1\rangle\otimes|1\rangle$.
\qed

\section{Proof of  Example~\ref{3^4}}
\label{sec:AppJ}

\noindent \p
$$\centering {\begin{array}{c c c}
\setlength{\arraycolsep}{2.0 pt}
\Psi=\begin{array}{lc}
\mbox{}&
\begin{array}{cccc}|0\rangle &|1\rangle  &|2\rangle &|3\rangle  \end{array}\\
\begin{array}{c}|0\rangle \\|1\rangle\\ |2\rangle\\ |3\rangle\\ \end{array}&
\begin{array}{|c|c|c|c|}
\hline
         &  |3\rangle  &|1\rangle& |2\rangle \\
\hline
|2\rangle &            &|3\rangle &|0\rangle \\
\hline
|3\rangle &|0\rangle   &         & |1\rangle  \\
\hline
|1\rangle & |2\rangle  &|0\rangle &          \\
\hline
\end{array}
\end{array}
\setlength{\arraycolsep}{2.4 pt}
~~~\Phi^1=\begin{array}{lc}
\mbox{}&
\begin{array}{ccc}|0\rangle &|1\rangle  &|2\rangle  \end{array}\\
\begin{array}{c}|0\rangle \\|1\rangle\\ |2\rangle\\ \end{array}&
\begin{array}{|c|c|c|}
\hline
|0\rangle & |1\rangle & |2\rangle \\
\hline
|1\rangle & |2\rangle & |0\rangle\\
\hline
|2\rangle & |0\rangle & |1\rangle \\
\hline
\end{array}
\end{array}
\setlength{\arraycolsep}{2.4 pt}
~~~\Phi^2=\begin{array}{lc}
\mbox{}&
\begin{array}{ccc}|0\rangle &|1\rangle  &|2\rangle  \end{array}\\
\begin{array}{c}|0\rangle \\|1\rangle\\ |2\rangle\\ \end{array}&
\begin{array}{|c|c|c|}
\hline
|0\rangle & |1\rangle & |2\rangle \\
\hline
|2\rangle & |0\rangle & |1\rangle\\
\hline
|1\rangle & |2\rangle & |0\rangle \\
\hline
\end{array}
\end{array}
\end{array}}$$
$$\setlength{\arraycolsep}{1.2 pt}
~~\Phi=\begin{array}{lc}
\mbox{}&
\begin{array}{cccccccccccc}~|0\rangle ~&|1\rangle~ & |2\rangle~&  ~|3\rangle&  ~|4\rangle ~& ~|5\rangle & ~|6\rangle & ~|7\rangle& ~|8\rangle & ~|9\rangle & |10\rangle & |11\rangle  \end{array}\\
\begin{array}{c}|0\rangle \\|1\rangle\\ |2\rangle\\ |3\rangle\\ |4\rangle\\ |5\rangle \\ |6\rangle \\ |7\rangle\\ |8\rangle \\ |9\rangle \\|10\rangle \\ |11\rangle \end{array}&
\begin{array}{|c|c|c|c|c|c|c|c|c|c|c|c|c|}
\hline
            &          &            &  |9\rangle & |10\rangle & |11\rangle & |3\rangle & |4\rangle & |5\rangle& |6\rangle & |7\rangle & |8\rangle\\
\hline
            &          &            &  |10\rangle & |11\rangle & |9\rangle & |4\rangle & |5\rangle & |3\rangle& |7\rangle & |8\rangle & |6\rangle\\
\hline
            &          &            &  |11\rangle & |9\rangle & |10\rangle & |4\rangle & |3\rangle & |4\rangle& |8\rangle & |6\rangle & |7\rangle\\
\hline
|6\rangle & |8\rangle &  |7\rangle  &            &            &            &|9\rangle & |10\rangle  &|11\rangle& |0\rangle& |1\rangle & |2\rangle\\
\hline
|7\rangle & |6\rangle &  |8\rangle  &            &            &            &|10\rangle & |11\rangle &|9\rangle & |1\rangle& |2\rangle & |0\rangle\\
\hline
|8\rangle & |7\rangle &  |6\rangle  &            &            &            &|11\rangle & |9\rangle  &|10\rangle& |2\rangle& |0\rangle & |1\rangle\\
\hline
|9\rangle &|11\rangle & |10\rangle&|0\rangle& |2\rangle & |1\rangle &             &            &           & |3\rangle  & |4\rangle & |5\rangle \\
\hline
|10\rangle&|9\rangle & |11\rangle&|1\rangle& |0\rangle & |2\rangle &             &            &           & |4\rangle  & |5\rangle & |3\rangle \\
\hline
|11\rangle&|10\rangle & |9\rangle&|2\rangle& |1\rangle & |0\rangle &             &            &           & |5\rangle  & |3\rangle & |4\rangle \\
\hline
|3\rangle&|5\rangle &|4\rangle &|6\rangle & |8\rangle &  |7\rangle &|0\rangle &|2\rangle &|1\rangle &           &             &      \\
\hline
|4\rangle&|3\rangle & |5\rangle &|7\rangle & |6\rangle &  |5\rangle &|1\rangle &|0\rangle &|2\rangle &           &             &      \\
\hline
|5\rangle&|4\rangle & |3\rangle &|8\rangle & |7\rangle &  |6\rangle &|2\rangle &|1\rangle &|0\rangle &           &             &      \\
\hline
\end{array}
\end{array}$$ \qed

\section{Proof of  Example~\ref{moqlcs16}}
\label{sec:AppK}
\noindent \p Let $\mathbb{C}^4=\Span \{|0\rangle,|1\rangle,|2\rangle,|3\rangle\}$. Then $\mathbb{C}^{16}\simeq \mathbb{C}^4\otimes \mathbb{C}^4 =\Span\{|i\rangle \otimes |j\rangle: i, j\in [4] \}=\Span\{|0\rangle,|1\rangle,\dots,|15\rangle\}$. Define $U=\sum\limits_{i\in [4]}|i\rangle\langle i|\otimes U_i$, where $U_i$, $i\in [4]$,  is the same as  in Eq.~(\ref{00}).

Let $L^j$ and $K^j$, $1\leq j\leq 3$, be the same triple of orthogonal classical quantum Latin cubes of dimension 4 as below.
$$
\begin{small}
\setlength{\arraycolsep}{2.0pt}L^1/K^1:
       \begin{array}{ccccccccccccccccccc}
|0\rangle&|1\rangle&|2\rangle&|3\rangle&~~&|1\rangle&|0\rangle&|3\rangle&|2\rangle&~~&|2\rangle&|3\rangle&|0\rangle&|1\rangle&~~&|3\rangle&|2\rangle&|1\rangle&|0\rangle\\
|1\rangle&|0\rangle&|3\rangle&|2\rangle&~~&|0\rangle&|1\rangle&|2\rangle&|3\rangle&~~&|3\rangle&|2\rangle&|1\rangle&|0\rangle&~~&|2\rangle&|3\rangle&|0\rangle&|1\rangle\\
|2\rangle&|3\rangle&|0\rangle&|1\rangle&~~&|3\rangle&|2\rangle&|1\rangle&|0\rangle&~~&|0\rangle&|1\rangle&|2\rangle&|3\rangle&~~&|1\rangle&|0\rangle&|3\rangle&|2\rangle\\
|3\rangle&|2\rangle&|1\rangle&|0\rangle&~~&|2\rangle&|3\rangle&|0\rangle&|1\rangle&~~&|1\rangle&|0\rangle&|3\rangle&|2\rangle&~~&|0\rangle&|1\rangle&|2\rangle&|3\rangle\\
\end{array}
\end{small}
$$
\hspace{3.4cm}$R^1$\hspace{1.8cm}$R^2$\hspace{2cm}$R^3$\hspace{2cm}$R^4$
$$
\begin{small}
\setlength{\arraycolsep}{2.0pt}L^2/K^2:
       \begin{array}{ccccccccccccccccccc}
|0\rangle&|3\rangle&|1\rangle&|2\rangle&~~&|1\rangle&|2\rangle&|0\rangle&|3\rangle&~~&|2\rangle&|1\rangle&|3\rangle&|0\rangle&~~&|3\rangle&|0\rangle&|2\rangle&|1\rangle\\
|2\rangle&|1\rangle&|3\rangle&|0\rangle&~~&|3\rangle&|0\rangle&|2\rangle&|1\rangle&~~&|0\rangle&|3\rangle&|1\rangle&|2\rangle&~~&|1\rangle&|2\rangle&|0\rangle&|3\rangle\\
|3\rangle&|0\rangle&|2\rangle&|1\rangle&~~&|2\rangle&|1\rangle&|3\rangle&|0\rangle&~~&|1\rangle&|2\rangle&|0\rangle&|3\rangle&~~&|0\rangle&|3\rangle&|1\rangle&|2\rangle\\
|1\rangle&|2\rangle&|0\rangle&|3\rangle&~~&|0\rangle&|3\rangle&|1\rangle&|2\rangle&~~&|3\rangle&|0\rangle&|2\rangle&|1\rangle&~~&|2\rangle&|1\rangle&|3\rangle&|0\rangle\\
\end{array}
\end{small}
$$
\hspace{3.4cm}$S^1$\hspace{1.8cm}$S^2$\hspace{2cm}$S^3$\hspace{2cm}$S^4$
$$
\begin{small}
\setlength{\arraycolsep}{2.0pt}L^3/K^3:
       \begin{array}{ccccccccccccccccccc}
|0\rangle&|2\rangle&|3\rangle&|1\rangle&~~&|1\rangle&|3\rangle&|2\rangle&|0\rangle&~~&|2\rangle&|0\rangle&|1\rangle&|3\rangle&~~&|3\rangle&|1\rangle&|0\rangle&|2\rangle\\
|3\rangle&|1\rangle&|0\rangle&|2\rangle&~~&|2\rangle&|0\rangle&|1\rangle&|3\rangle&~~&|1\rangle&|3\rangle&|2\rangle&|0\rangle&~~&|0\rangle&|2\rangle&|3\rangle&|1\rangle\\
|1\rangle&|3\rangle&|2\rangle&|0\rangle&~~&|0\rangle&|2\rangle&|3\rangle&|1\rangle&~~&|3\rangle&|1\rangle&|0\rangle&|2\rangle&~~&|2\rangle&|0\rangle&|1\rangle&|3\rangle\\
|2\rangle&|0\rangle&|1\rangle&|3\rangle&~~&|3\rangle&|1\rangle&|0\rangle&|2\rangle&~~&|0\rangle&|2\rangle&|3\rangle&|1\rangle&~~&|1\rangle&|3\rangle&|2\rangle&|0\rangle\\
\end{array}
\end{small}
$$
\hspace{3.4cm}$T^1$\hspace{1.8cm}$T^2$\hspace{2cm}$T^3$\hspace{2cm}$T^4$

\noindent where $R^i$, $S^i$ and $T^i$, $1\leq i \leq4$, are respectively the four planes of classical quantum Latin cubes $L^1$, $L^2$ and $L^3$ or $K^1$, $K^2$ and $K^3$ in one direction,  the remaining eight planes from other two directions also can be drawn from $L^j$ or $K^j$, $1\leq j\leq 3$.  Then the following $\Phi$, $\Psi$ and $\Upsilon$ is a set of 3-MOQLC(16).
$$
\begin{small}
\setlength{\arraycolsep}{2.8pt}\Phi:
       \begin{array}{cccccccccccc}
|0\rangle \otimes R^i&|1\rangle\otimes R^i&|2\rangle\otimes R^i&|3\rangle\otimes R^i&~~&|1\rangle\otimes R^i&|0\rangle\otimes R^i&|3\rangle\otimes R^i&|2\rangle\otimes R^i&\\
|1\rangle\otimes R^i&|0\rangle\otimes R^i&|3\rangle\otimes U_3 R^i&|2\rangle\otimes R^i&~~&|0\rangle\otimes R^i&|1\rangle\otimes R^i&|2\rangle\otimes R^i&|3\rangle\otimes R^i&~\\
|2\rangle\otimes R^i&|3\rangle\otimes R^i&|0\rangle\otimes R^i&|1\rangle\otimes R^i&~~&|3\rangle\otimes R^i&|2\rangle\otimes U_2 R^i&|1\rangle\otimes R^i&|0\rangle\otimes R^i&~\\
|3\rangle\otimes R^i&|2\rangle\otimes R^i&|1\rangle\otimes R^i&|0\rangle\otimes R^i&~~&|2\rangle\otimes R^i&|3\rangle\otimes R^i&|0\rangle\otimes R^i&|1\rangle\otimes R^i&~\\
\end{array}
\end{small}
$$
\hspace{3.1cm}$\Phi^{i1}$\hspace{5.7cm}$\Phi^{i2}$
$$
\begin{small}
\setlength{\arraycolsep}{2.0pt}
  \begin{array}{cccccccccccc}
|2\rangle\otimes R^i&|3\rangle\otimes R^i&|0\rangle\otimes R^i&|1\rangle\otimes R^i&~~&|3\rangle\otimes U_3 R^i&|2\rangle\otimes R^i&|1\rangle\otimes R^i&|0\rangle\otimes R^i\\
|3\rangle\otimes R^i&|2\rangle\otimes R^i&|1\rangle\otimes U_1 R^i&|0\rangle\otimes R^i&~~&|2\rangle\otimes R^i&|3\rangle\otimes R^i&|0\rangle\otimes R^i&|1\rangle\otimes R^i\\
|0\rangle\otimes R^i&|1\rangle\otimes R^i&|2\rangle\otimes R^i&|3\rangle\otimes R^i&~~&|1\rangle\otimes R^i&|0\rangle\otimes R^i&|3\rangle\otimes R^i&|2\rangle\otimes R^i\\
|1\rangle\otimes R^i&|0\rangle\otimes R^i&|3\rangle\otimes R^i&|2\rangle\otimes R^i&~~&|0\rangle\otimes R^i&|1\rangle\otimes R^i&|2\rangle\otimes R^i&|3\rangle\otimes R^i
\end{array}
\end{small}
$$
\hspace{3.1cm}$\Phi^{i3}$\hspace{5.7cm}$\Phi^{i4}$
$$
\begin{small}
\setlength{\arraycolsep}{2.8pt}\Psi:
       \begin{array}{ccccccccccccc}
|0\rangle\otimes S^i&|3\rangle\otimes S^i&|1\rangle\otimes S^i&|2\rangle\otimes S^i&~~&|1\rangle\otimes S^i&|2\rangle\otimes S^i&|0\rangle\otimes S^i&|3\rangle\otimes S^i&~&\\
|2\rangle\otimes S^i&|1\rangle\otimes U_1 S^i&|3\rangle\otimes S^i&|0\rangle\otimes S^i&~~&|3\rangle\otimes S^i&|0\rangle\otimes S^i&|2\rangle\otimes S^i&|1\rangle\otimes S^i&~&\\
|3\rangle\otimes S^i&|0\rangle\otimes S^i&|2\rangle\otimes S^i&|1\rangle\otimes S^i&~~&|2\rangle\otimes S^i&|1\rangle\otimes S^i&|3\rangle\otimes U_3 S^i&|0\rangle\otimes S^i&~&\\
|1\rangle\otimes S^i&|2\rangle\otimes S^i&|0\rangle\otimes S^i&|3\rangle\otimes S^i&~~&|0\rangle\otimes S^i&|3\rangle\otimes S^i&|1\rangle\otimes S^i&|2\rangle\otimes S^i&~&\\
\end{array}
\end{small}
$$
\hspace{3.1cm}$\Psi^{i1}$\hspace{5.7cm}$\Psi^{i2}$
$$
\begin{small}
\setlength{\arraycolsep}{2.8pt}
       \begin{array}{ccccccccccccc}
|2\rangle\otimes S^i&|1\rangle\otimes S^i&|3\rangle\otimes S^i&|0\rangle\otimes S^i&~~&|3\rangle\otimes S^i&|0\rangle\otimes S^i&|2\rangle\otimes U_2 S^i&|1\rangle\otimes S^i\\
|0\rangle\otimes S^i&|3\rangle\otimes S^i&|1\rangle\otimes S^i&|2\rangle\otimes S^i&~~&|1\rangle\otimes S^i&|2\rangle\otimes S^i&|0\rangle\otimes S^i&|3\rangle\otimes S^i\\
|1\rangle\otimes S^i&|2\rangle\otimes S^i&|0\rangle\otimes U_0 S^i&|3\rangle\otimes S^i&~~&|0\rangle\otimes S^i&|3\rangle\otimes S^i&|1\rangle\otimes S^i&|2\rangle\otimes S^i\\
|3\rangle\otimes S^i&|0\rangle\otimes S^i&|2\rangle\otimes S^i&|1\rangle\otimes S^i&~~&|2\rangle\otimes S^i&|1\rangle\otimes S^i&|3\rangle\otimes S^i&|0\rangle\otimes S^i
\end{array}
\end{small}
$$
\hspace{3.1cm}$\Psi^{i3}$\hspace{5.7cm}$\Psi^{i4}$
$$
\begin{small}
\setlength{\arraycolsep}{2.8pt}\Upsilon:
       \begin{array}{ccccccccccccccccccc}
|0\rangle\otimes T^i&|2\rangle\otimes U_2 T^i&|3\rangle\otimes T^i&|1\rangle\otimes T^i&~~&|1\rangle\otimes T^i&|3\rangle\otimes T^i&|2\rangle\otimes T^i&|0\rangle\otimes T^i&~&\\
|3\rangle\otimes T^i&|1\rangle\otimes T^i&|0\rangle\otimes T^i&|2\rangle\otimes T^i&~~&|2\rangle\otimes T^i&|0\rangle\otimes T^i&|1\rangle\otimes T^i&|3\rangle\otimes T^i&~&\\
|1\rangle\otimes T^i&|3\rangle\otimes T^i&|2\rangle\otimes T^i&|0\rangle\otimes T^i&~~&|0\rangle\otimes T^i&|2\rangle\otimes U_2 T^i&|3\rangle\otimes T^i&|1\rangle\otimes T^i&~&\\
|2\rangle\otimes T^i&|0\rangle\otimes T^i&|1\rangle\otimes T^i&|3\rangle\otimes T^i&~~&|3\rangle\otimes T^i&|1\rangle\otimes T^i&|0\rangle\otimes T^i&|2\rangle\otimes T^i&~&\\
\end{array}
\end{small}
$$
\hspace{3.1cm}$\Upsilon^{i1}$\hspace{5.7cm}$\Upsilon^{i2}$\newpage
$$
\begin{small}
\setlength{\arraycolsep}{2.8pt}
       \begin{array}{ccccccccccccccccccc}
|2\rangle\otimes T^i&|0\rangle\otimes T^i&|1\rangle\otimes T^i&|3\rangle\otimes T^i&~~&|3\rangle\otimes T^i&|1\rangle\otimes T^i&|0\rangle\otimes T^i&|2\rangle\otimes T^i\\
|1\rangle\otimes T^i&|3\rangle\otimes T^i&|2\rangle\otimes T^i&|0\rangle\otimes T^i&~~&|0\rangle\otimes T^i&|2\rangle\otimes T^i&|3\rangle\otimes T^i&|1\rangle\otimes T^i\\
|3\rangle\otimes T^i&|1\rangle\otimes T^i&|0\rangle\otimes U_0 T^i&|2\rangle\otimes T^i&~~&|2\rangle\otimes T^i&|0\rangle\otimes T^i&|1\rangle\otimes T^i&|3\rangle\otimes T^i\\
|0\rangle\otimes T^i&|2\rangle\otimes T^i&|3\rangle\otimes T^i&|1\rangle\otimes T^i&~~&|1\rangle\otimes T^i&|3\rangle\otimes U_3 T^i&|2\rangle\otimes T^i&|0\rangle\otimes T^i
\end{array}
\end{small}
$$
\hspace{3.1cm}$\Upsilon^{i3}$\hspace{5.7cm}$\Upsilon^{i4}$

\noindent Here $\Phi^{il}$, $\Psi^{il}$ and $\Upsilon^{il}$, $1 \leq i,l \leq 4$, are respectively the sixteen planes of $\Phi$, $\Psi$ and $\Upsilon$ in one direction.
And in some small blocks, such as $|j\rangle \otimes U_jR^{i}$, $U_jR^{i}$ means that $U_j$ acts on every elements of $R^i$.

Further, $\Phi$, $\Psi$ and $\Upsilon$ are  non-classical quantum Latin cubes.  For $\Phi$, let us take $|3\rangle \otimes U_3R^1$ and $|3\rangle\otimes R^1$ in the plane $\Phi^{11}$.
$$\begin{small}\setlength{\arraycolsep}{2.0pt}  {\begin{array}{|c|c|c|c|}
\hline
|3\rangle \otimes U_3|0\rangle& |3\rangle \otimes U_3|1\rangle &  |3\rangle \otimes U_3|2\rangle&  |3\rangle \otimes U_3|3\rangle \\
\hline
 |3\rangle \otimes U_3|1\rangle &  |3\rangle \otimes U_3|0\rangle& |3\rangle \otimes  U_3|3\rangle&  |3\rangle \otimes U_3|2\rangle \\
\hline
 |3\rangle \otimes U_3|2\rangle &  |3\rangle \otimes U_3|3\rangle &  |3\rangle \otimes U_3|0\rangle & |3\rangle \otimes  U_3|1\rangle \\
\hline
 |3\rangle \otimes U_3|3\rangle & |3\rangle \otimes  U_3|2\rangle &  |3\rangle \otimes U_3|1\rangle &  |3\rangle \otimes U_3|0\rangle \\
\hline
\end{array}} ~ ~~ {\begin{array}{|c|c|c|c|}
\hline
 |3\rangle \otimes |0\rangle& |3\rangle \otimes  |1\rangle & |3\rangle \otimes  |2\rangle & |3\rangle \otimes  |3\rangle \\
\hline
 |3\rangle \otimes |1\rangle &  |3\rangle \otimes |0\rangle & |3\rangle \otimes  |3\rangle& |3\rangle \otimes  |2\rangle \\
\hline
 |3\rangle \otimes |2\rangle &  |3\rangle \otimes |3\rangle& |3\rangle \otimes  |0\rangle& |3\rangle \otimes  |1\rangle \\
\hline
 |3\rangle \otimes |3\rangle& |3\rangle \otimes  |2\rangle& |3\rangle \otimes  |1\rangle&  |3\rangle \otimes |0\rangle \\
\hline
\end{array}}
\end{small}$$
For instance, in the 1st row, 3rd column of the left table and the 1st row, 2nd column of the right table, we have $|\langle3|3\rangle\langle 2|U^{\dagger}_3 |1\rangle|=|\frac{\sqrt{6}}{4}|\neq 0$ or $\neq 1$. Thus $\Phi$ is a non-classical quantum Latin cube. The same holds for
$\Psi$ and $\Upsilon$.
\qed

\end{document}